\documentstyle [twoside,12pt] {article}
\catcode`@=11 \catcode`!=11

\expandafter\ifx\csname fiverm\endcsname\relax
  
\fi

\let\!latexendpicture=\endpicture
\let\!latexframe=\frame
\let\!latexlinethickness=\linethickness
\let\!latexmultiput=\multiput
\let\!latexput=\put

\def\@picture(#1,#2)(#3,#4){%
  \@picht #2\unitlength
  \setbox\@picbox\hbox to #1\unitlength\bgroup
  \let\endpicture=\!latexendpicture
  \let\frame=\!latexframe
  \let\linethickness=\!latexlinethickness
  \let\multiput=\!latexmultiput
  \let\put=\!latexput
  \hskip -#3\unitlength \lower #4\unitlength \hbox\bgroup}

\catcode`@=12 \catcode`!=12

\catcode`!=11 

\def\PiC{P\kern-.12em\lower.5ex\hbox{I}\kern-.075emC}
\def\PiCTeX{\PiC\kern-.11em\TeX}

\def\!ifnextchar#1#2#3{%
  \let\!testchar=#1%
  \def\!first{#2}%
  \def\!second{#3}%
  \futurelet\!nextchar\!testnext}
\def\!testnext{%
  \ifx \!nextchar \!spacetoken
    \let\!next=\!skipspacetestagain
  \else
    \ifx \!nextchar \!testchar
      \let\!next=\!first
    \else
      \let\!next=\!second
    \fi
  \fi
  \!next}
\def\\{\!skipspacetestagain}
  \expandafter\def\\ {\futurelet\!nextchar\!testnext}
\def\\{\let\!spacetoken= } \\  

\def\!tfor#1:=#2\do#3{%
  \edef\!fortemp{#2}%
  \ifx\!fortemp\!empty
    \else
    \!tforloop#2\!nil\!nil\!!#1{#3}%
  \fi}
\def\!tforloop#1#2\!!#3#4{%
  \def#3{#1}%
  \ifx #3\!nnil
    \let\!nextwhile=\!fornoop
  \else
    #4\relax
    \let\!nextwhile=\!tforloop
  \fi
  \!nextwhile#2\!!#3{#4}}

\def\!etfor#1:=#2\do#3{%
  \def\!!tfor{\!tfor#1:=}%
  \edef\!!!tfor{#2}%
  \expandafter\!!tfor\!!!tfor\do{#3}}

\def\!cfor#1:=#2\do#3{%
  \edef\!fortemp{#2}%
  \ifx\!fortemp\!empty
  \else
    \!cforloop#2,\!nil,\!nil\!!#1{#3}%
  \fi}
\def\!cforloop#1,#2\!!#3#4{%
  \def#3{#1}%
  \ifx #3\!nnil
    \let\!nextwhile=\!fornoop
  \else
    #4\relax
    \let\!nextwhile=\!cforloop
  \fi
  \!nextwhile#2\!!#3{#4}}

\def\!ecfor#1:=#2\do#3{%
  \def\!!cfor{\!cfor#1:=}%
  \edef\!!!cfor{#2}%
  \expandafter\!!cfor\!!!cfor\do{#3}}

\def\!empty{}
\def\!nnil{\!nil}
\def\!fornoop#1\!!#2#3{}

\def\!ifempty#1#2#3{%
  \edef\!emptyarg{#1}%
  \ifx\!emptyarg\!empty
    #2%
  \else
    #3%
  \fi}

\def\!getnext#1\from#2{%
  \expandafter\!gnext#2\!#1#2}%
\def\!gnext\\#1#2\!#3#4{%
  \def#3{#1}%
  \def#4{#2\\{#1}}%
  \ignorespaces}

\def\!getnextvalueof#1\from#2{%
  \expandafter\!gnextv#2\!#1#2}%
\def\!gnextv\\#1#2\!#3#4{%
  #3=#1%
  \def#4{#2\\{#1}}%
  \ignorespaces}

\def\!copylist#1\to#2{%
  \expandafter\!!copylist#1\!#2}
\def\!!copylist#1\!#2{%
  \def#2{#1}\ignorespaces}

\def\!wlet#1=#2{%
  \let#1=#2
  \wlog{\string#1=\string#2}}

\def\!listaddon#1#2{%
  \expandafter\!!listaddon#2\!{#1}#2}
\def\!!listaddon#1\!#2#3{%
  \def#3{#1\\#2}}

\def\!rightappend#1\withCS#2\to#3{\expandafter\!!rightappend#3\!#2{#1}#3}
\def\!!rightappend#1\!#2#3#4{\def#4{#1#2{#3}}}

\def\!leftappend#1\withCS#2\to#3{\expandafter\!!leftappend#3\!#2{#1}#3}
\def\!!leftappend#1\!#2#3#4{\def#4{#2{#3}#1}}

\def\!lop#1\to#2{\expandafter\!!lop#1\!#1#2}
\def\!!lop\\#1#2\!#3#4{\def#4{#1}\def#3{#2}}

\def\!loop#1\repeat{\def\!body{#1}\!iterate}
\def\!iterate{\!body\let\!next=\!iterate\else\let\!next=\relax\fi\!next}

\def\!!loop#1\repeat{\def\!!body{#1}\!!iterate}
\def\!!iterate{\!!body\let\!!next=\!!iterate\else\let\!!next=\relax\fi\!!next}

\def\!removept#1#2{\edef#2{\expandafter\!!removePT\the#1}}
{\catcode`p=12 \catcode`t=12 \gdef\!!removePT#1pt{#1}}

\def\placevalueinpts of <#1> in #2 {%
  \!removept{#1}{#2}}

\def\!mlap#1{\hbox to 0pt{\hss#1\hss}}
\def\!vmlap#1{\vbox to 0pt{\vss#1\vss}}

\def\!not#1{%
  #1\relax
    \!switchfalse
  \else
    \!switchtrue
  \fi
  \if!switch
  \ignorespaces}

\let\!!!wlog=\wlog              
\def\wlog#1{}

\newdimen\headingtoplotskip     
\newdimen\linethickness         
\newdimen\longticklength        
\newdimen\plotsymbolspacing     
\newdimen\shortticklength       
\newdimen\stackleading          
\newdimen\tickstovaluesleading  
\newdimen\totalarclength        
\newdimen\valuestolabelleading  

\newbox\!boxA                   
\newbox\!boxB                   
\newbox\!picbox                 
\newbox\!plotsymbol             
\newbox\!putobject              
\newbox\!shadesymbol            

\newcount\!countA               
\newcount\!countB               
\newcount\!countC               
\newcount\!countD               
\newcount\!countE               
\newcount\!countF               
\newcount\!countG               
\newcount\!fiftypt              
\newcount\!intervalno           
\newcount\!npoints              
\newcount\!nsegments            
\newcount\!ntemp                
\newcount\!parity               
\newcount\!scalefactor          
\newcount\!tfs                  
\newcount\!tickcase             

\newdimen\!Xleft                
\newdimen\!Xright               
\newdimen\!Xsave                
\newdimen\!Ybot                 
\newdimen\!Ysave                
\newdimen\!Ytop                 
\newdimen\!angle                
\newdimen\!arclength            
\newdimen\!areabloc             
\newdimen\!arealloc             
\newdimen\!arearloc             
\newdimen\!areatloc             
\newdimen\!bshrinkage           
\newdimen\!checkbot             
\newdimen\!checkleft            
\newdimen\!checkright           
\newdimen\!checktop             
\newdimen\!dimenA               
\newdimen\!dimenB               
\newdimen\!dimenC               
\newdimen\!dimenD               
\newdimen\!dimenE               
\newdimen\!dimenF               
\newdimen\!dimenG               
\newdimen\!dimenH               
\newdimen\!dimenI               
\newdimen\!distacross           
\newdimen\!downlength           
\newdimen\!dp                   
\newdimen\!dshade               
\newdimen\!dxpos                
\newdimen\!dxprime              
\newdimen\!dypos                
\newdimen\!dyprime              
\newdimen\!ht                   
\newdimen\!leaderlength         
\newdimen\!lshrinkage           
\newdimen\!midarclength         
\newdimen\!offset               
\newdimen\!plotheadingoffset    
\newdimen\!plotsymbolxshift     
\newdimen\!plotsymbolyshift     
\newdimen\!plotxorigin          
\newdimen\!plotyorigin          
\newdimen\!rootten              
\newdimen\!rshrinkage           
\newdimen\!shadesymbolxshift    
\newdimen\!shadesymbolyshift    
\newdimen\!tenAa                
\newdimen\!tenAc                
\newdimen\!tenAe                
\newdimen\!tshrinkage           
\newdimen\!uplength             
\newdimen\!wd                   
\newdimen\!wmax                 
\newdimen\!wmin                 
\newdimen\!xB                   
\newdimen\!xC                   
\newdimen\!xE                   
\newdimen\!xM                   
\newdimen\!xS                   
\newdimen\!xaxislength          
\newdimen\!xdiff                
\newdimen\!xleft                
\newdimen\!xloc                 
\newdimen\!xorigin              
\newdimen\!xpivot               
\newdimen\!xpos                 
\newdimen\!xprime               
\newdimen\!xright               
\newdimen\!xshade               
\newdimen\!xshift               
\newdimen\!xtemp                
\newdimen\!xunit                
\newdimen\!xxE                  
\newdimen\!xxM                  
\newdimen\!xxS                  
\newdimen\!xxloc                
\newdimen\!yB                   
\newdimen\!yC                   
\newdimen\!yE                   
\newdimen\!yM                   
\newdimen\!yS                   
\newdimen\!yaxislength          
\newdimen\!ybot                 
\newdimen\!ydiff                
\newdimen\!yloc                 
\newdimen\!yorigin              
\newdimen\!ypivot               
\newdimen\!ypos                 
\newdimen\!yprime               
\newdimen\!yshade               
\newdimen\!yshift               
\newdimen\!ytemp                
\newdimen\!ytop                 
\newdimen\!yunit                
\newdimen\!yyE                  
\newdimen\!yyM                  
\newdimen\!yyS                  
\newdimen\!yyloc                
\newdimen\!zpt                  

\newif\if!axisvisible           
\newif\if!gridlinestoo          
\newif\if!keepPO                
\newif\if!placeaxislabel        
\newif\if!switch                
\newif\if!xswitch               

\newtoks\!axisLaBeL             
\newtoks\!keywordtoks           

\newwrite\!replotfile           

\newhelp\!keywordhelp{The keyword mentioned in the error message in unknown.
Replace NEW KEYWORD in the indicated response by the keyword that
should have been specified.}    

\!wlet\!!origin=\!xM                   
\!wlet\!!unit=\!uplength               
\!wlet\!Lresiduallength=\!dimenG       
\!wlet\!Rresiduallength=\!dimenF       
\!wlet\!axisLength=\!distacross        
\!wlet\!axisend=\!ydiff                
\!wlet\!axisstart=\!xdiff              
\!wlet\!axisxlevel=\!arclength         
\!wlet\!axisylevel=\!downlength        
\!wlet\!beta=\!dimenE                  
\!wlet\!gamma=\!dimenF                 
\!wlet\!shadexorigin=\!plotxorigin     
\!wlet\!shadeyorigin=\!plotyorigin     
\!wlet\!ticklength=\!xS                
\!wlet\!ticklocation=\!xE              
\!wlet\!ticklocationincr=\!yE          
\!wlet\!tickwidth=\!yS                 
\!wlet\!totalleaderlength=\!dimenE     
\!wlet\!xone=\!xprime                  
\!wlet\!xtwo=\!dxprime                 
\!wlet\!ySsave=\!yM                    
\!wlet\!ybB=\!yB                       
\!wlet\!ybC=\!yC                       
\!wlet\!ybE=\!yE                       
\!wlet\!ybM=\!yM                       
\!wlet\!ybS=\!yS                       
\!wlet\!ybpos=\!yyloc                  
\!wlet\!yone=\!yprime                  
\!wlet\!ytB=\!xB                       
\!wlet\!ytC=\!xC                       
\!wlet\!ytE=\!downlength               
\!wlet\!ytM=\!arclength                
\!wlet\!ytS=\!distacross               
\!wlet\!ytpos=\!xxloc                  
\!wlet\!ytwo=\!dyprime                 

\!zpt=0pt                              
\!xunit=1pt
\!yunit=1pt
\!arearloc=\!xunit
\!areatloc=\!yunit
\!dshade=5pt
\!leaderlength=24in
\!tfs=256                              
\!wmax=5.3pt                           
\!wmin=2.7pt                           
\!xaxislength=\!xunit
\!xpivot=\!zpt
\!yaxislength=\!yunit
\!ypivot=\!zpt
  \!dimenA=50pt \!fiftypt=\!dimenA     

\!rootten=3.162278pt                   
\!tenAa=8.690286pt                     
\!tenAc=2.773839pt                     
\!tenAe=2.543275pt                     

\def\!cosrotationangle{1}      
\def\!sinrotationangle{0}      
\def\!xpivotcoord{0}           
\def\!xref{0}                  
\def\!xshadesave{0}            
\def\!ypivotcoord{0}           
\def\!yref{0}                  
\def\!yshadesave{0}            
\def\!zero{0}                  

\let\wlog=\!!!wlog
\def\normalgraphs{%
  \longticklength=.4\baselineskip
  \shortticklength=.25\baselineskip
  \tickstovaluesleading=.25\baselineskip
  \valuestolabelleading=.8\baselineskip
  \linethickness=.4pt
  \stackleading=.17\baselineskip
  \headingtoplotskip=1.5\baselineskip
  \visibleaxes
  \ticksout
  \nogridlines
  \unloggedticks}
\def\setplotarea x from #1 to #2, y from #3 to #4 {%
  \!arealloc=\!M{#1}\!xunit \advance \!arealloc -\!xorigin
  \!areabloc=\!M{#3}\!yunit \advance \!areabloc -\!yorigin
  \!arearloc=\!M{#2}\!xunit \advance \!arearloc -\!xorigin
  \!areatloc=\!M{#4}\!yunit \advance \!areatloc -\!yorigin
  \!initinboundscheck
  \!xaxislength=\!arearloc  \advance\!xaxislength -\!arealloc
  \!yaxislength=\!areatloc  \advance\!yaxislength -\!areabloc
  \!plotheadingoffset=\!zpt
  \!dimenput {{\setbox0=\hbox{}\wd0=\!xaxislength\ht0=\!yaxislength\box0}}
     [bl] (\!arealloc,\!areabloc)}
\def\visibleaxes{%
  \def\!axisvisibility{\!axisvisibletrue}}

\def\!fixkeyword#1{%
  \errhelp=\!keywordhelp
  \errmessage{Unrecognized keyword `#1': \the\!keywordtoks{NEW KEYWORD}'}}

\!keywordtoks={enter `i\fixkeyword}

\def\fixkeyword#1{%
  \!nextkeyword#1 }

\def\axis {%
  \def\!nextkeyword##1 {%
    \expandafter\ifx\csname !axis##1\endcsname \relax
      \def\!next{\!fixkeyword{##1}}%
    \else
      \def\!next{\csname !axis##1\endcsname}%
    \fi
    \!next}%
  \!offset=\!zpt
  \!axisvisibility
  \!placeaxislabelfalse
  \!nextkeyword}

\def\!axisbottom{%
  \!axisylevel=\!areabloc
  \def\!tickxsign{0}%
  \def\!tickysign{-}%
  \def\!axissetup{\!axisxsetup}%
  \def\!axislabeltbrl{t}%
  \!nextkeyword}

\def\!axistop{%
  \!axisylevel=\!areatloc
  \def\!tickxsign{0}%
  \def\!tickysign{+}%
  \def\!axissetup{\!axisxsetup}%
  \def\!axislabeltbrl{b}%
  \!nextkeyword}

\def\!axisleft{%
  \!axisxlevel=\!arealloc
  \def\!tickxsign{-}%
  \def\!tickysign{0}%
  \def\!axissetup{\!axisysetup}%
  \def\!axislabeltbrl{r}%
  \!nextkeyword}

\def\!axisright{%
  \!axisxlevel=\!arearloc
  \def\!tickxsign{+}%
  \def\!tickysign{0}%
  \def\!axissetup{\!axisysetup}%
  \def\!axislabeltbrl{l}%
  \!nextkeyword}

\def\!axisshiftedto#1=#2 {%
  \if 0\!tickxsign
    \!axisylevel=\!M{#2}\!yunit
    \advance\!axisylevel -\!yorigin
  \else
    \!axisxlevel=\!M{#2}\!xunit
    \advance\!axisxlevel -\!xorigin
  \fi
  \!nextkeyword}

\def\!axisvisible{%
  \!axisvisibletrue
  \!nextkeyword}

\def\!axisinvisible{%
  \!axisvisiblefalse
  \!nextkeyword}

\def\!axislabel#1 {%
  \!axisLaBeL={#1}%
  \!placeaxislabeltrue
  \!nextkeyword}

\expandafter\def\csname !axis/\endcsname{%
  \!axissetup 
  \if!placeaxislabel
    \!placeaxislabel
  \fi
  \if +\!tickysign 
    \!dimenA=\!axisylevel
    \advance\!dimenA \!offset 
    \advance\!dimenA -\!areatloc 
    \ifdim \!dimenA>\!plotheadingoffset
      \!plotheadingoffset=\!dimenA 
    \fi
  \fi}

\def\grid #1 #2 {%
  \!countA=#1\advance\!countA 1
  \axis bottom invisible ticks length <\!zpt> andacross quantity {\!countA} /
  \!countA=#2\advance\!countA 1
  \axis left   invisible ticks length <\!zpt> andacross quantity {\!countA} / }

\def\plotheading#1 {%
  \advance\!plotheadingoffset \headingtoplotskip
  \!dimenput {#1} [B] <.5\!xaxislength,\!plotheadingoffset>
    (\!arealloc,\!areatloc)}

\def\!axisxsetup{%
  \!axisxlevel=\!arealloc
  \!axisstart=\!arealloc
  \!axisend=\!arearloc
  \!axisLength=\!xaxislength
  \!!origin=\!xorigin
  \!!unit=\!xunit
  \!xswitchtrue
  \if!axisvisible
    \!makeaxis
  \fi}

\def\!axisysetup{%
  \!axisylevel=\!areabloc
  \!axisstart=\!areabloc
  \!axisend=\!areatloc
  \!axisLength=\!yaxislength
  \!!origin=\!yorigin
  \!!unit=\!yunit
  \!xswitchfalse
  \if!axisvisible
    \!makeaxis
  \fi}

\def\!makeaxis{%
  \setbox\!boxA=\hbox{
    \beginpicture
      \!setdimenmode
      \setcoordinatesystem point at {\!zpt} {\!zpt}
      \putrule from {\!zpt} {\!zpt} to
        {\!tickysign\!tickysign\!axisLength}
        {\!tickxsign\!tickxsign\!axisLength}
    \endpicturesave <\!Xsave,\!Ysave>}%
    \wd\!boxA=\!zpt
    \!placetick\!axisstart}

\def\!placeaxislabel{%
  \advance\!offset \valuestolabelleading
  \if!xswitch
    \!dimenput {\the\!axisLaBeL} [\!axislabeltbrl]
      <.5\!axisLength,\!tickysign\!offset> (\!axisxlevel,\!axisylevel)
    \advance\!offset \!dp  
    \advance\!offset \!ht  
  \else
    \!dimenput {\the\!axisLaBeL} [\!axislabeltbrl]
      <\!tickxsign\!offset,.5\!axisLength> (\!axisxlevel,\!axisylevel)
  \fi
  \!axisLaBeL={}}

\def\arrow <#1> [#2,#3]{%
  \!ifnextchar<{\!arrow{#1}{#2}{#3}}{\!arrow{#1}{#2}{#3}<\!zpt,\!zpt> }}

\def\!arrow#1#2#3<#4,#5> from #6 #7 to #8 #9 {%
  \!xloc=\!M{#8}\!xunit
  \!yloc=\!M{#9}\!yunit
  \!dxpos=\!xloc  \!dimenA=\!M{#6}\!xunit  \advance \!dxpos -\!dimenA
  \!dypos=\!yloc  \!dimenA=\!M{#7}\!yunit  \advance \!dypos -\!dimenA
  \let\!MAH=\!M
  \!setdimenmode
  \!xshift=#4\relax  \!yshift=#5\relax
  \!reverserotateonly\!xshift\!yshift
  \advance\!xshift\!xloc  \advance\!yshift\!yloc
  \!xS=-\!dxpos  \advance\!xS\!xshift
  \!yS=-\!dypos  \advance\!yS\!yshift
  \!start (\!xS,\!yS)
  \!ljoin (\!xshift,\!yshift)
  \!Pythag\!dxpos\!dypos\!arclength
  \!divide\!dxpos\!arclength\!dxpos
  \!dxpos=32\!dxpos  \!removept\!dxpos\!!cos
  \!divide\!dypos\!arclength\!dypos
  \!dypos=32\!dypos  \!removept\!dypos\!!sin
  \!halfhead{#1}{#2}{#3}
  \!halfhead{#1}{-#2}{-#3}
  \let\!M=\!MAH
  \ignorespaces}
  \def\!halfhead#1#2#3{%
    \!dimenC=-#1%
    \divide \!dimenC 2 
    \!dimenD=#2\!dimenC
    \!rotate(\!dimenC,\!dimenD)by(\!!cos,\!!sin)to(\!xM,\!yM)
    \!dimenC=-#1
    \!dimenD=#3\!dimenC
    \!dimenD=.5\!dimenD
    \!rotate(\!dimenC,\!dimenD)by(\!!cos,\!!sin)to(\!xE,\!yE)
    \!start (\!xshift,\!yshift)
    \advance\!xM\!xshift  \advance\!yM\!yshift
    \advance\!xE\!xshift  \advance\!yE\!yshift
    \!qjoin (\!xM,\!yM) (\!xE,\!yE)
    \ignorespaces}

\def\betweenarrows #1#2 from #3 #4 to #5 #6 {%
  \!xloc=\!M{#3}\!xunit  \!xxloc=\!M{#5}\!xunit%
  \!yloc=\!M{#4}\!yunit  \!yyloc=\!M{#6}\!yunit%
  \!dxpos=\!xxloc  \advance\!dxpos by -\!xloc
  \!dypos=\!yyloc  \advance\!dypos by -\!yloc
  \advance\!xloc .5\!dxpos
  \advance\!yloc .5\!dypos
  \let\!MBA=\!M
  \!setdimenmode
  \ifdim\!dypos=\!zpt
    \ifdim\!dxpos<\!zpt \!dxpos=-\!dxpos \fi
    \put {\!lrarrows{\!dxpos}{#1}}#2{} at {\!xloc} {\!yloc}
  \else
    \ifdim\!dxpos=\!zpt
      \ifdim\!dypos<\!zpt \!dypos=-\!zpt \fi
      \put {\!udarrows{\!dypos}{#1}}#2{} at {\!xloc} {\!yloc}
    \fi
  \fi
  \let\!M=\!MBA
  \ignorespaces}

\def\!lrarrows#1#2{
  {\setbox\!boxA=\hbox{$\mkern-2mu\mathord-\mkern-2mu$}%
   \setbox\!boxB=\hbox{$\leftarrow$}\!dimenE=\ht\!boxB
   \setbox\!boxB=\hbox{}\ht\!boxB=2\!dimenE
   \hbox to #1{$\mathord\leftarrow\mkern-6mu
     \cleaders\copy\!boxA\hfil
     \mkern-6mu\mathord-$%
     \kern.4em $\vcenter{\box\!boxB}$$\vcenter{\hbox{#2}}$\kern.4em
     $\mathord-\mkern-6mu
     \cleaders\copy\!boxA\hfil
     \mkern-6mu\mathord\rightarrow$}}}

\def\!udarrows#1#2{
  {\setbox\!boxB=\hbox{#2}%
   \setbox\!boxA=\hbox to \wd\!boxB{\hss$\vert$\hss}%
   \!dimenE=\ht\!boxA \advance\!dimenE \dp\!boxA \divide\!dimenE 2
   \vbox to #1{\offinterlineskip
      \vskip .05556\!dimenE
      \hbox to \wd\!boxB{\hss$\mkern.4mu\uparrow$\hss}\vskip-\!dimenE
      \cleaders\copy\!boxA\vfil
      \vskip-\!dimenE\copy\!boxA
      \vskip\!dimenE\copy\!boxB\vskip.4em
      \copy\!boxA\vskip-\!dimenE
      \cleaders\copy\!boxA\vfil
      \vskip-\!dimenE \hbox to \wd\!boxB{\hss$\mkern.4mu\downarrow$\hss}
      \vskip .05556\!dimenE}}}

\def\putbar#1breadth <#2> from #3 #4 to #5 #6 {%
  \!xloc=\!M{#3}\!xunit  \!xxloc=\!M{#5}\!xunit%
  \!yloc=\!M{#4}\!yunit  \!yyloc=\!M{#6}\!yunit%
  \!dypos=\!yyloc  \advance\!dypos by -\!yloc
  \!dimenI=#2
  \ifdim \!dimenI=\!zpt 
    \putrule#1from {#3} {#4} to {#5} {#6} 
  \else 
    \let\!MBar=\!M
    \!setdimenmode 
    \divide\!dimenI 2
    \ifdim \!dypos=\!zpt
      \advance \!yloc -\!dimenI 
      \advance \!yyloc \!dimenI
    \else
      \advance \!xloc -\!dimenI 
      \advance \!xxloc \!dimenI
    \fi
    \putrectangle#1corners at {\!xloc} {\!yloc} and {\!xxloc} {\!yyloc}
    \let\!M=\!MBar 
  \fi
  \ignorespaces}

\def\setbars#1breadth <#2> baseline at #3 = #4 {%
  \edef\!barshift{#1}%
  \edef\!barbreadth{#2}%
  \edef\!barorientation{#3}%
  \edef\!barbaseline{#4}%
  \def\!bardobaselabel{\!bardoendlabel}%
  \def\!bardoendlabel{\!barfinish}%
  \let\!drawcurve=\!barcurve
  \!setbars}
\def\!setbars{%
  \futurelet\!nextchar\!!setbars}
\def\!!setbars{%
  \if b\!nextchar
    \def\!!!setbars{\!setbarsbget}%
  \else
    \if e\!nextchar
      \def\!!!setbars{\!setbarseget}%
    \else
      \def\!!!setbars{\relax}%
    \fi
  \fi
  \!!!setbars}
\def\!setbarsbget baselabels (#1) {%
  \def\!barbaselabelorientation{#1}%
  \def\!bardobaselabel{\!!bardobaselabel}%
  \!setbars}
\def\!setbarseget endlabels (#1) {%
  \edef\!barendlabelorientation{#1}%
  \def\!bardoendlabel{\!!bardoendlabel}%
  \!setbars}

\def\!barcurve #1 #2 {%
  \if y\!barorientation
    \def\!basexarg{#1}%
    \def\!baseyarg{\!barbaseline}%
  \else
    \def\!basexarg{\!barbaseline}%
    \def\!baseyarg{#2}%
  \fi
  \expandafter\putbar\!barshift breadth <\!barbreadth> from {\!basexarg}
    {\!baseyarg} to {#1} {#2}
  \def\!endxarg{#1}%
  \def\!endyarg{#2}%
  \!bardobaselabel}

\def\!!bardobaselabel "#1" {%
  \put {#1}\!barbaselabelorientation{} at {\!basexarg} {\!baseyarg}
  \!bardoendlabel}

\def\!!bardoendlabel "#1" {%
  \put {#1}\!barendlabelorientation{} at {\!endxarg} {\!endyarg}
  \!barfinish}

\def\!barfinish{%
  \!ifnextchar/{\!finish}{\!barcurve}}

\def\putrectangle{%
  \!ifnextchar<{\!putrectangle}{\!putrectangle<\!zpt,\!zpt> }}
\def\!putrectangle<#1,#2> corners at #3 #4 and #5 #6 {%
  \!xone=\!M{#3}\!xunit  \!xtwo=\!M{#5}\!xunit%
  \!yone=\!M{#4}\!yunit  \!ytwo=\!M{#6}\!yunit%
  \ifdim \!xtwo<\!xone
    \!dimenI=\!xone  \!xone=\!xtwo  \!xtwo=\!dimenI
  \fi
  \ifdim \!ytwo<\!yone
    \!dimenI=\!yone  \!yone=\!ytwo  \!ytwo=\!dimenI
  \fi
  \!dimenI=#1\relax  \advance\!xone\!dimenI  \advance\!xtwo\!dimenI
  \!dimenI=#2\relax  \advance\!yone\!dimenI  \advance\!ytwo\!dimenI
  \let\!MRect=\!M
  \!setdimenmode
  \!shaderectangle
  \!dimenI=.5\linethickness
  \advance \!xone  -\!dimenI
  \advance \!xtwo   \!dimenI
  \putrule from {\!xone} {\!yone} to {\!xtwo} {\!yone}
  \putrule from {\!xone} {\!ytwo} to {\!xtwo} {\!ytwo}
  \advance \!xone   \!dimenI
  \advance \!xtwo  -\!dimenI%
  \advance \!yone  -\!dimenI
  \advance \!ytwo   \!dimenI
  \putrule from {\!xone} {\!yone} to {\!xone} {\!ytwo}
  \putrule from {\!xtwo} {\!yone} to {\!xtwo} {\!ytwo}
  \let\!M=\!MRect
  \ignorespaces}

\def\shaderectanglesoff{%
  \def\!shaderectangle{}%
  \ignorespaces}

\shaderectanglesoff

\def\!!shaderectangle{%
  \!dimenA=\!xtwo  \advance \!dimenA -\!xone
  \!dimenB=\!ytwo  \advance \!dimenB -\!yone
  \ifdim \!dimenA<\!dimenB
    \!startvshade (\!xone,\!yone,\!ytwo)
    \!lshade      (\!xtwo,\!yone,\!ytwo)
  \else
    \!starthshade (\!yone,\!xone,\!xtwo)
    \!lshade      (\!ytwo,\!xone,\!xtwo)
  \fi
  \ignorespaces}

\def\frame{%
  \!ifnextchar<{\!frame}{\!frame<\!zpt> }}
\long\def\!frame<#1> #2{%
  \beginpicture
    \setcoordinatesystem units <1pt,1pt> point at 0 0
    \put {#2} [Bl] at 0 0
    \!dimenA=#1\relax
    \!dimenB=\!wd \advance \!dimenB \!dimenA
    \!dimenC=\!ht \advance \!dimenC \!dimenA
    \!dimenD=\!dp \advance \!dimenD \!dimenA
    \let\!MFr=\!M
    \!setdimenmode
    \putrectangle corners at {-\!dimenA} {-\!dimenD} and {\!dimenB} {\!dimenC}
    \!setcoordmode
    \let\!M=\!MFr
  \endpicture
  \ignorespaces}

\def\rectangle <#1> <#2> {%
  \setbox0=\hbox{}\wd0=#1\ht0=#2\frame {\box0}}

\def\plot{%
  \!ifnextchar"{\!plotfromfile}{\!drawcurve}}
\def\!plotfromfile"#1"{%
  \expandafter\!drawcurve \input #1 /}

\def\setquadratic{%
  \let\!drawcurve=\!qcurve
  \let\!!Shade=\!!qShade
  \let\!!!Shade=\!!!qShade}

\def\setlinear{%
  \let\!drawcurve=\!lcurve
  \let\!!Shade=\!!lShade
  \let\!!!Shade=\!!!lShade}

\def\sethistograms{%
  \let\!drawcurve=\!hcurve}

\def\!qcurve #1 #2 {%
  \!start (#1,#2)
  \!Qjoin}
\def\!Qjoin#1 #2 #3 #4 {%
  \!qjoin (#1,#2) (#3,#4)             
  \!ifnextchar/{\!finish}{\!Qjoin}}

\def\!lcurve #1 #2 {%
  \!start (#1,#2)
  \!Ljoin}
\def\!Ljoin#1 #2 {%
  \!ljoin (#1,#2)                    
  \!ifnextchar/{\!finish}{\!Ljoin}}

\def\!finish/{\ignorespaces}

\def\!hcurve #1 #2 {%
  \edef\!hxS{#1}%
  \edef\!hyS{#2}%
  \!hjoin}
\def\!hjoin#1 #2 {%
  \putrectangle corners at {\!hxS} {\!hyS} and {#1} {#2}
  \edef\!hxS{#1}%
  \!ifnextchar/{\!finish}{\!hjoin}}

\def\vshade #1 #2 #3 {%
  \!startvshade (#1,#2,#3)
  \!Shadewhat}

\def\hshade #1 #2 #3 {%
  \!starthshade (#1,#2,#3)
  \!Shadewhat}

\def\!Shadewhat{%
  \futurelet\!nextchar\!Shade}
\def\!Shade{%
  \if <\!nextchar
    \def\!nextShade{\!!Shade}%
  \else
    \if /\!nextchar
      \def\!nextShade{\!finish}%
    \else
      \def\!nextShade{\!!!Shade}%
    \fi
  \fi
  \!nextShade}
\def\!!lShade<#1> #2 #3 #4 {%
  \!lshade <#1> (#2,#3,#4)                 
  \!Shadewhat}
\def\!!!lShade#1 #2 #3 {%
  \!lshade (#1,#2,#3)
  \!Shadewhat}
\def\!!qShade<#1> #2 #3 #4 #5 #6 #7 {%
  \!qshade <#1> (#2,#3,#4) (#5,#6,#7)      
  \!Shadewhat}
\def\!!!qShade#1 #2 #3 #4 #5 #6 {%
  \!qshade (#1,#2,#3) (#4,#5,#6)
  \!Shadewhat}

\setlinear

\def\setdashpattern <#1>{%
  \def\!Flist{}\def\!Blist{}\def\!UDlist{}%
  \!countA=0
  \!ecfor\!item:=#1\do{%
    \!dimenA=\!item\relax
    \expandafter\!rightappend\the\!dimenA\withCS{\\}\to\!UDlist%
    \advance\!countA  1
    \ifodd\!countA
      \expandafter\!rightappend\the\!dimenA\withCS{\!Rule}\to\!Flist%
      \expandafter\!leftappend\the\!dimenA\withCS{\!Rule}\to\!Blist%
    \else
      \expandafter\!rightappend\the\!dimenA\withCS{\!Skip}\to\!Flist%
      \expandafter\!leftappend\the\!dimenA\withCS{\!Skip}\to\!Blist%
    \fi}%
  \!leaderlength=\!zpt
  \def\!Rule##1{\advance\!leaderlength  ##1}%
  \def\!Skip##1{\advance\!leaderlength  ##1}%
  \!Flist%
  \ifdim\!leaderlength>\!zpt
  \else
    \def\!Flist{\!Skip{24in}}\def\!Blist{\!Skip{24in}}\ignorespaces
    \def\!UDlist{\\{\!zpt}\\{24in}}\ignorespaces
    \!leaderlength=24in
  \fi
  \!dashingon}

\def\!dashingon{%
  \def\!advancedashing{\!!advancedashing}%
  \def\!drawlinearsegment{\!lineardashed}%
  \def\!puthline{\!putdashedhline}%
  \def\!putvline{\!putdashedvline}%
  \ignorespaces}%
\def\!dashingoff{%
  \def\!advancedashing{\relax}%
  \def\!drawlinearsegment{\!linearsolid}%
  \def\!puthline{\!putsolidhline}%
  \def\!putvline{\!putsolidvline}%
  \ignorespaces}

\def\setdots{%
  \!ifnextchar<{\!setdots}{\!setdots<5pt>}}
\def\!setdots<#1>{%
  \!dimenB=#1\advance\!dimenB -\plotsymbolspacing
  \ifdim\!dimenB<\!zpt
    \!dimenB=\!zpt
  \fi
\setdashpattern <\plotsymbolspacing,\!dimenB>}

\def\setdotsnear <#1> for <#2>{%
  \!dimenB=#2\relax  \advance\!dimenB -.05pt
  \!dimenC=#1\relax  \!countA=\!dimenC
  \!dimenD=\!dimenB  \advance\!dimenD .5\!dimenC  \!countB=\!dimenD
  \divide \!countB  \!countA
  \ifnum 1>\!countB
    \!countB=1
  \fi
  \divide\!dimenB  \!countB
  \setdots <\!dimenB>}

\def\setdashes{%
  \!ifnextchar<{\!setdashes}{\!setdashes<5pt>}}
\def\!setdashes<#1>{\setdashpattern <#1,#1>}

\def\setdashesnear <#1> for <#2>{%
  \!dimenB=#2\relax
  \!dimenC=#1\relax  \!countA=\!dimenC
  \!dimenD=\!dimenB  \advance\!dimenD .5\!dimenC  \!countB=\!dimenD
  \divide \!countB  \!countA
  \ifodd \!countB
  \else
    \advance \!countB  1
  \fi
  \divide\!dimenB  \!countB
  \setdashes <\!dimenB>}

\def\setsolid{%
  \def\!Flist{\!Rule{24in}}\def\!Blist{\!Rule{24in}}%
  \def\!UDlist{\\{24in}\\{\!zpt}}%
  \!dashingoff}
\setsolid

\def\!divide#1#2#3{%
  \!dimenB=#1
  \!dimenC=#2
  \!dimenD=\!dimenB
  \divide \!dimenD \!dimenC
  \!dimenA=\!dimenD
  \multiply\!dimenD \!dimenC
  \advance\!dimenB -\!dimenD
  \!dimenD=\!dimenC
    \ifdim\!dimenD<\!zpt \!dimenD=-\!dimenD
  \fi
  \ifdim\!dimenD<64pt
    \!divstep[\!tfs]\!divstep[\!tfs]%
  \else
    \!!divide
  \fi
  #3=\!dimenA\ignorespaces}

\def\!!divide{%
  \ifdim\!dimenD<256pt
    \!divstep[64]\!divstep[32]\!divstep[32]%
  \else
    \!divstep[8]\!divstep[8]\!divstep[8]\!divstep[8]\!divstep[8]%
    \!dimenA=2\!dimenA
  \fi}

\def\!divstep[#1]{
  \!dimenB=#1\!dimenB
  \!dimenD=\!dimenB
    \divide \!dimenD by \!dimenC
  \!dimenA=#1\!dimenA
    \advance\!dimenA by \!dimenD%
  \multiply\!dimenD by \!dimenC
    \advance\!dimenB by -\!dimenD}

\def\Divide <#1> by <#2> forming <#3> {%
  \!divide{#1}{#2}{#3}}

\def\circulararc{%
  \ellipticalarc axes ratio 1:1 }

\def\ellipticalarc axes ratio #1:#2 #3 degrees from #4 #5 center at #6 #7 {%
  \!angle=#3pt\relax
  \ifdim\!angle>\!zpt
    \def\!sign{}
  \else
    \def\!sign{-}\!angle=-\!angle
  \fi
  \!xxloc=\!M{#6}\!xunit
  \!yyloc=\!M{#7}\!yunit
  \!xxS=\!M{#4}\!xunit
  \!yyS=\!M{#5}\!yunit
  \advance\!xxS -\!xxloc
  \advance\!yyS -\!yyloc
  \!divide\!xxS{#1pt}\!xxS 
  \!divide\!yyS{#2pt}\!yyS 
  \let\!MC=\!M
  \!setdimenmode
  \!xS=#1\!xxS  \advance\!xS\!xxloc
  \!yS=#2\!yyS  \advance\!yS\!yyloc
  \!start (\!xS,\!yS)%
  \!loop\ifdim\!angle>14.9999pt
    \!rotate(\!xxS,\!yyS)by(\!cos,\!sign\!sin)to(\!xxM,\!yyM)
    \!rotate(\!xxM,\!yyM)by(\!cos,\!sign\!sin)to(\!xxE,\!yyE)
    \!xM=#1\!xxM  \advance\!xM\!xxloc  \!yM=#2\!yyM  \advance\!yM\!yyloc
    \!xE=#1\!xxE  \advance\!xE\!xxloc  \!yE=#2\!yyE  \advance\!yE\!yyloc
    \!qjoin (\!xM,\!yM) (\!xE,\!yE)
    \!xxS=\!xxE  \!yyS=\!yyE
    \advance \!angle -15pt
  \repeat
  \ifdim\!angle>\!zpt
    \!angle=100.53096\!angle
    \divide \!angle 360 
    \!sinandcos\!angle\!!sin\!!cos
    \!rotate(\!xxS,\!yyS)by(\!!cos,\!sign\!!sin)to(\!xxM,\!yyM)
    \!rotate(\!xxM,\!yyM)by(\!!cos,\!sign\!!sin)to(\!xxE,\!yyE)
    \!xM=#1\!xxM  \advance\!xM\!xxloc  \!yM=#2\!yyM  \advance\!yM\!yyloc
    \!xE=#1\!xxE  \advance\!xE\!xxloc  \!yE=#2\!yyE  \advance\!yE\!yyloc
    \!qjoin (\!xM,\!yM) (\!xE,\!yE)
  \fi
  \let\!M=\!MC
  \ignorespaces}

\def\!rotate(#1,#2)by(#3,#4)to(#5,#6){%
  \!dimenA=#3#1\advance \!dimenA -#4#2
  \!dimenB=#3#2\advance \!dimenB  #4#1
  \divide \!dimenA 32  \divide \!dimenB 32
  #5=\!dimenA  #6=\!dimenB
  \ignorespaces}
\def\!sin{4.17684}
\def\!cos{31.72624}

\def\!sinandcos#1#2#3{%
 \!dimenD=#1
 \!dimenA=\!dimenD
 \!dimenB=32pt
 \!removept\!dimenD\!value
 \!dimenC=\!dimenD
 \!dimenC=\!value\!dimenC \divide\!dimenC by 64 
 \advance\!dimenB by -\!dimenC
 \!dimenC=\!value\!dimenC \divide\!dimenC by 96 
 \advance\!dimenA by -\!dimenC
 \!dimenC=\!value\!dimenC \divide\!dimenC by 128 
 \advance\!dimenB by \!dimenC%
 \!removept\!dimenA#2
 \!removept\!dimenB#3
 \ignorespaces}

\def\putrule#1from #2 #3 to #4 #5 {%
  \!xloc=\!M{#2}\!xunit  \!xxloc=\!M{#4}\!xunit%
  \!yloc=\!M{#3}\!yunit  \!yyloc=\!M{#5}\!yunit%
  \!dxpos=\!xxloc  \advance\!dxpos by -\!xloc
  \!dypos=\!yyloc  \advance\!dypos by -\!yloc
  \ifdim\!dypos=\!zpt
    \def\!!Line{\!puthline{#1}}\ignorespaces
  \else
    \ifdim\!dxpos=\!zpt
      \def\!!Line{\!putvline{#1}}\ignorespaces
    \else
       \def\!!Line{}
    \fi
  \fi
  \let\!ML=\!M
  \!setdimenmode
  \!!Line%
  \let\!M=\!ML
  \ignorespaces}

\def\!putsolidhline#1{%
  \ifdim\!dxpos>\!zpt
    \put{\!hline\!dxpos}#1[l] at {\!xloc} {\!yloc}
  \else
    \put{\!hline{-\!dxpos}}#1[l] at {\!xxloc} {\!yyloc}
  \fi
  \ignorespaces}

\def\!putsolidvline#1{%
  \ifdim\!dypos>\!zpt
    \put{\!vline\!dypos}#1[b] at {\!xloc} {\!yloc}
  \else
    \put{\!vline{-\!dypos}}#1[b] at {\!xxloc} {\!yyloc}
  \fi
  \ignorespaces}

\def\!hline#1{\hbox to #1{\leaders \hrule height\linethickness\hfill}}
\def\!vline#1{\vbox to #1{\leaders \vrule width\linethickness\vfill}}

\def\!putdashedhline#1{%
  \ifdim\!dxpos>\!zpt
    \!DLsetup\!Flist\!dxpos
    \put{\hbox to \!totalleaderlength{\!hleaders}\!hpartialpattern\!Rtrunc}
      #1[l] at {\!xloc} {\!yloc}
  \else
    \!DLsetup\!Blist{-\!dxpos}
    \put{\!hpartialpattern\!Ltrunc\hbox to \!totalleaderlength{\!hleaders}}
      #1[r] at {\!xloc} {\!yloc}
  \fi
  \ignorespaces}

\def\!putdashedvline#1{%
  \!dypos=-\!dypos
  \ifdim\!dypos>\!zpt
    \!DLsetup\!Flist\!dypos
    \put{\vbox{\vbox to \!totalleaderlength{\!vleaders}
      \!vpartialpattern\!Rtrunc}}#1[t] at {\!xloc} {\!yloc}
  \else
    \!DLsetup\!Blist{-\!dypos}
    \put{\vbox{\!vpartialpattern\!Ltrunc
      \vbox to \!totalleaderlength{\!vleaders}}}#1[b] at {\!xloc} {\!yloc}
  \fi
  \ignorespaces}

\def\!DLsetup#1#2{
  \let\!RSlist=#1
  \!countB=#2
  \!countA=\!leaderlength
  \divide\!countB by \!countA
  \!totalleaderlength=\!countB\!leaderlength
  \!Rresiduallength=#2%
  \advance \!Rresiduallength by -\!totalleaderlength
  \!Lresiduallength=\!leaderlength
  \advance \!Lresiduallength by -\!Rresiduallength
  \ignorespaces}

\def\!hleaders{%
  \def\!Rule##1{\vrule height\linethickness width##1}%
  \def\!Skip##1{\hskip##1}%
  \leaders\hbox{\!RSlist}\hfill}

\def\!hpartialpattern#1{%
  \!dimenA=\!zpt \!dimenB=\!zpt
  \def\!Rule##1{#1{##1}\vrule height\linethickness width\!dimenD}%
  \def\!Skip##1{#1{##1}\hskip\!dimenD}%
  \!RSlist}

\def\!vleaders{%
  \def\!Rule##1{\hrule width\linethickness height##1}%
  \def\!Skip##1{\vskip##1}%
  \leaders\vbox{\!RSlist}\vfill}

\def\!vpartialpattern#1{%
  \!dimenA=\!zpt \!dimenB=\!zpt
  \def\!Rule##1{#1{##1}\hrule width\linethickness height\!dimenD}%
  \def\!Skip##1{#1{##1}\vskip\!dimenD}%
  \!RSlist}

\def\!Rtrunc#1{\!trunc{#1}>\!Rresiduallength}
\def\!Ltrunc#1{\!trunc{#1}<\!Lresiduallength}

\def\!trunc#1#2#3{%
  \!dimenA=\!dimenB
  \advance\!dimenB by #1%
  \!dimenD=\!dimenB  \ifdim\!dimenD#2#3\!dimenD=#3\fi
  \!dimenC=\!dimenA  \ifdim\!dimenC#2#3\!dimenC=#3\fi
  \advance \!dimenD by -\!dimenC}

\def\!start (#1,#2){%
  \!plotxorigin=\!xorigin  \advance \!plotxorigin by \!plotsymbolxshift
  \!plotyorigin=\!yorigin  \advance \!plotyorigin by \!plotsymbolyshift
  \!xS=\!M{#1}\!xunit \!yS=\!M{#2}\!yunit
  \!rotateaboutpivot\!xS\!yS
  \!copylist\!UDlist\to\!!UDlist
  \!getnextvalueof\!downlength\from\!!UDlist
  \!distacross=\!zpt
  \!intervalno=0 
  \global\totalarclength=\!zpt
  \ignorespaces}

\def\!ljoin (#1,#2){%
  \advance\!intervalno by 1
  \!xE=\!M{#1}\!xunit \!yE=\!M{#2}\!yunit
  \!rotateaboutpivot\!xE\!yE
  \!xdiff=\!xE \advance \!xdiff by -\!xS
  \!ydiff=\!yE \advance \!ydiff by -\!yS
  \!Pythag\!xdiff\!ydiff\!arclength
  \global\advance \totalarclength by \!arclength%
  \!drawlinearsegment
  \!xS=\!xE \!yS=\!yE
  \ignorespaces}

\def\!linearsolid{%
  \!npoints=\!arclength
  \!countA=\plotsymbolspacing
  \divide\!npoints by \!countA
  \ifnum \!npoints<1
    \!npoints=1
  \fi
  \divide\!xdiff by \!npoints
  \divide\!ydiff by \!npoints
  \!xpos=\!xS \!ypos=\!yS
  \loop\ifnum\!npoints>-1
    \!plotifinbounds
    \advance \!xpos by \!xdiff
    \advance \!ypos by \!ydiff
    \advance \!npoints by -1
  \repeat
  \ignorespaces}

\def\!lineardashed{%
  \ifdim\!distacross>\!arclength
    \advance \!distacross by -\!arclength  
  \else
    \loop\ifdim\!distacross<\!arclength
      \!divide\!distacross\!arclength\!dimenA
      \!removept\!dimenA\!t
      \!xpos=\!t\!xdiff \advance \!xpos by \!xS
      \!ypos=\!t\!ydiff \advance \!ypos by \!yS
      \!plotifinbounds
      \advance\!distacross by \plotsymbolspacing
      \!advancedashing
    \repeat
    \advance \!distacross by -\!arclength
  \fi
  \ignorespaces}

\def\!!advancedashing{%
  \advance\!downlength by -\plotsymbolspacing
  \ifdim \!downlength>\!zpt
  \else
    \advance\!distacross by \!downlength
    \!getnextvalueof\!uplength\from\!!UDlist
    \advance\!distacross by \!uplength
    \!getnextvalueof\!downlength\from\!!UDlist
  \fi}

\def\inboundscheckoff{%
  \def\!plotifinbounds{\!plot(\!xpos,\!ypos)}%
  \def\!initinboundscheck{\relax}\ignorespaces}

\inboundscheckoff

\def\!!plotifinbounds{%
  \ifdim \!xpos<\!checkleft
  \else
    \ifdim \!xpos>\!checkright
    \else
      \ifdim \!ypos<\!checkbot
      \else
         \ifdim \!ypos>\!checktop
         \else
           \!plot(\!xpos,\!ypos)
         \fi
      \fi
    \fi
  \fi}

\def\!!initinboundscheck{%
  \!checkleft=\!arealloc     \advance\!checkleft by \!xorigin
  \!checkright=\!arearloc    \advance\!checkright by \!xorigin
  \!checkbot=\!areabloc      \advance\!checkbot by \!yorigin
  \!checktop=\!areatloc      \advance\!checktop by \!yorigin}

\def\!logten#1#2{%
  \expandafter\!!logten#1\!nil
  \!removept\!dimenF#2%
  \ignorespaces}

\def\!!logten#1#2\!nil{%
  \if -#1%
    \!dimenF=\!zpt
    \def\!next{\ignorespaces}%
  \else
    \if +#1%
      \def\!next{\!!logten#2\!nil}%
    \else
      \if .#1%
        \def\!next{\!!logten0.#2\!nil}%
      \else
        \def\!next{\!!!logten#1#2..\!nil}%
      \fi
    \fi
  \fi
  \!next}

\def\!!!logten#1#2.#3.#4\!nil{%
  \!dimenF=1pt 
  \if 0#1%
    \!!logshift#3pt 
  \else 
    \!logshift#2/
    \!dimenE=#1.#2#3pt 
  \fi 
  \ifdim \!dimenE<\!rootten
    \multiply \!dimenE 10 
    \advance  \!dimenF -1pt
  \fi
  \!dimenG=\!dimenE
    \advance\!dimenG 10pt
  \advance\!dimenE -10pt 
  \multiply\!dimenE 10 
  \!divide\!dimenE\!dimenG\!dimenE
  \!removept\!dimenE\!t
  \!dimenG=\!t\!dimenE
  \!removept\!dimenG\!tt
  \!dimenH=\!tt\!tenAe
    \divide\!dimenH 100
  \advance\!dimenH \!tenAc
  \!dimenH=\!tt\!dimenH
    \divide\!dimenH 100
  \advance\!dimenH \!tenAa
  \!dimenH=\!t\!dimenH
    \divide\!dimenH 100 
  \advance\!dimenF \!dimenH}

\def\!logshift#1{%
  \if #1/%
    \def\!next{\ignorespaces}%
  \else
    \advance\!dimenF 1pt
    \def\!next{\!logshift}%
  \fi
  \!next}

 \def\!!logshift#1{%
   \advance\!dimenF -1pt
   \if 0#1%
     \def\!next{\!!logshift}%
   \else
     \if p#1%
       \!dimenF=1pt
       \def\!next{\!dimenE=1p}%
     \else
       \def\!next{\!dimenE=#1.}%
     \fi
   \fi
   \!next}

\def\beginpicture{%
  \setbox\!picbox=\hbox\bgroup%
  \!xleft=\maxdimen
  \!xright=-\maxdimen
  \!ybot=\maxdimen
  \!ytop=-\maxdimen}

\def\endpicture{%
  \ifdim\!xleft=\maxdimen
    \!xleft=\!zpt \!xright=\!zpt \!ybot=\!zpt \!ytop=\!zpt
  \fi
  \global\!Xleft=\!xleft \global\!Xright=\!xright
  \global\!Ybot=\!ybot \global\!Ytop=\!ytop
  \egroup%
  \ht\!picbox=\!Ytop  \dp\!picbox=-\!Ybot
  \ifdim\!Ybot>\!zpt
  \else
    \ifdim\!Ytop<\!zpt
      \!Ybot=\!Ytop
    \else
      \!Ybot=\!zpt
    \fi
  \fi
  \hbox{\kern-\!Xleft\lower\!Ybot\box\!picbox\kern\!Xright}}

\def\endpicturesave <#1,#2>{%
  \endpicture \global #1=\!Xleft \global #2=\!Ybot \ignorespaces}

\def\setcoordinatesystem{%
  \!ifnextchar{u}{\!getlengths }
    {\!getlengths units <\!xunit,\!yunit>}}
\def\!getlengths units <#1,#2>{%
  \!xunit=#1\relax
  \!yunit=#2\relax
  \!ifcoordmode
    \let\!SCnext=\!SCccheckforRP
  \else
    \let\!SCnext=\!SCdcheckforRP
  \fi
  \!SCnext}
\def\!SCccheckforRP{%
  \!ifnextchar{p}{\!cgetreference }
    {\!cgetreference point at {\!xref} {\!yref} }}
\def\!cgetreference point at #1 #2 {%
  \edef\!xref{#1}\edef\!yref{#2}%
  \!xorigin=\!xref\!xunit  \!yorigin=\!yref\!yunit
  \!initinboundscheck 
  \ignorespaces}
\def\!SCdcheckforRP{%
  \!ifnextchar{p}{\!dgetreference}%
    {\ignorespaces}}
\def\!dgetreference point at #1 #2 {%
  \!xorigin=#1\relax  \!yorigin=#2\relax
  \ignorespaces}

\long\def\put#1#2 at #3 #4 {%
  \!setputobject{#1}{#2}%
  \!xpos=\!M{#3}\!xunit  \!ypos=\!M{#4}\!yunit
  \!rotateaboutpivot\!xpos\!ypos%
  \advance\!xpos -\!xorigin  \advance\!xpos -\!xshift
  \advance\!ypos -\!yorigin  \advance\!ypos -\!yshift
  \kern\!xpos\raise\!ypos\box\!putobject\kern-\!xpos%
  \!doaccounting\ignorespaces}

\long\def\multiput #1#2 at {%
  \!setputobject{#1}{#2}%
  \!ifnextchar"{\!putfromfile}{\!multiput}}
\def\!putfromfile"#1"{%
  \expandafter\!multiput \input #1 /}
\def\!multiput{%
  \futurelet\!nextchar\!!multiput}
\def\!!multiput{%
  \if *\!nextchar
    \def\!nextput{\!alsoby}%
  \else
    \if /\!nextchar
      \def\!nextput{\!finishmultiput}%
    \else
      \def\!nextput{\!alsoat}%
    \fi
  \fi
  \!nextput}
\def\!finishmultiput/{%
  \setbox\!putobject=\hbox{}%
  \ignorespaces}

\def\!alsoat#1 #2 {%
  \!xpos=\!M{#1}\!xunit  \!ypos=\!M{#2}\!yunit
  \!rotateaboutpivot\!xpos\!ypos%
  \advance\!xpos -\!xorigin  \advance\!xpos -\!xshift
  \advance\!ypos -\!yorigin  \advance\!ypos -\!yshift
  \kern\!xpos\raise\!ypos\copy\!putobject\kern-\!xpos%
  \!doaccounting
  \!multiput}

\def\!alsoby*#1 #2 #3 {%
  \!dxpos=\!M{#2}\!xunit \!dypos=\!M{#3}\!yunit
  \!rotateonly\!dxpos\!dypos
  \!ntemp=#1%
  \!!loop\ifnum\!ntemp>0
    \advance\!xpos by \!dxpos  \advance\!ypos by \!dypos
    \kern\!xpos\raise\!ypos\copy\!putobject\kern-\!xpos%
    \advance\!ntemp by -1
  \repeat
  \!doaccounting
  \!multiput}

\def\accountingon{\def\!doaccounting{\!!doaccounting}\ignorespaces}

\accountingon
\def\!!doaccounting{%
  \!xtemp=\!xpos
  \!ytemp=\!ypos
  \ifdim\!xtemp<\!xleft
     \!xleft=\!xtemp
  \fi
  \advance\!xtemp by  \!wd
  \ifdim\!xright<\!xtemp
    \!xright=\!xtemp
  \fi
  \advance\!ytemp by -\!dp
  \ifdim\!ytemp<\!ybot
    \!ybot=\!ytemp
  \fi
  \advance\!ytemp by  \!dp
  \advance\!ytemp by  \!ht
  \ifdim\!ytemp>\!ytop
    \!ytop=\!ytemp
  \fi}

\long\def\!setputobject#1#2{%
  \setbox\!putobject=\hbox{#1}%
  \!ht=\ht\!putobject  \!dp=\dp\!putobject  \!wd=\wd\!putobject
  \wd\!putobject=\!zpt
  \!xshift=.5\!wd   \!yshift=.5\!ht   \advance\!yshift by -.5\!dp
  \edef\!putorientation{#2}%
  \expandafter\!SPOreadA\!putorientation[]\!nil%
  \expandafter\!SPOreadB\!putorientation<\!zpt,\!zpt>\!nil\ignorespaces}

\def\!SPOreadA#1[#2]#3\!nil{\!etfor\!orientation:=#2\do\!SPOreviseshift}

\def\!SPOreadB#1<#2,#3>#4\!nil{\advance\!xshift by -#2\advance\!yshift by -#3}

\def\!SPOreviseshift{%
  \if l\!orientation
    \!xshift=\!zpt
  \else
    \if r\!orientation
      \!xshift=\!wd
    \else
      \if b\!orientation
        \!yshift=-\!dp
      \else
        \if B\!orientation
          \!yshift=\!zpt
        \else
          \if t\!orientation
            \!yshift=\!ht
          \fi
        \fi
      \fi
    \fi
  \fi}

\long\def\!dimenput#1#2(#3,#4){%
  \!setputobject{#1}{#2}%
  \!xpos=#3\advance\!xpos by -\!xshift
  \!ypos=#4\advance\!ypos by -\!yshift
  \kern\!xpos\raise\!ypos\box\!putobject\kern-\!xpos%
  \!doaccounting\ignorespaces}

\def\!setdimenmode{%
  \let\!M=\!M!!\ignorespaces}
\def\!setcoordmode{%
  \let\!M=\!M!\ignorespaces}
\def\!ifcoordmode{%
  \ifx \!M \!M!}
\def\!ifdimenmode{%
  \ifx \!M \!M!!}
\def\!M!#1#2{#1#2}
\def\!M!!#1#2{#1}
\!setcoordmode
\let\setdimensionmode=\!setdimenmode
\let\setcoordinatemode=\!setcoordmode

\def\!stack[#1]{%
  \let\!lglue=\hfill \let\!rglue=\hfill
  \expandafter\let\csname !#1glue\endcsname=\relax
  \!ifnextchar<{\!!stack}{\!!stack<\stackleading>}}
\def\!!stack<#1>#2{%
  \vbox{\def\!valueslist{}\!ecfor\!value:=#2\do{%
    \expandafter\!rightappend\!value\withCS{\\}\to\!valueslist}%
    \!lop\!valueslist\to\!value
    \let\\=\cr\lineskiplimit=\maxdimen\lineskip=#1%
    \baselineskip=-1000pt\halign{\!lglue##\!rglue\cr \!value\!valueslist\cr}}%
  \ignorespaces}

\def\!lines[#1]#2{%
  \let\!lglue=\hfill \let\!rglue=\hfill
  \expandafter\let\csname !#1glue\endcsname=\relax
  \vbox{\halign{\!lglue##\!rglue\cr #2\crcr}}%
  \ignorespaces}

\def\!Lines[#1]#2{%
  \let\!lglue=\hfill \let\!rglue=\hfill
  \expandafter\let\csname !#1glue\endcsname=\relax
  \vtop{\halign{\!lglue##\!rglue\cr #2\crcr}}%
  \ignorespaces}

\def\setplotsymbol(#1#2){%
  \!setputobject{#1}{#2}
  \setbox\!plotsymbol=\box\!putobject%
  \!plotsymbolxshift=\!xshift
  \!plotsymbolyshift=\!yshift
  \ignorespaces}

\setplotsymbol({\rm\vpt .})

\def\!!plot(#1,#2){%
  \!dimenA=-\!plotxorigin \advance \!dimenA by #1
  \!dimenB=-\!plotyorigin \advance \!dimenB by #2
  \kern\!dimenA\raise\!dimenB\copy\!plotsymbol\kern-\!dimenA%
  \ignorespaces}

\def\!!!plot(#1,#2){%
  \!dimenA=-\!plotxorigin \advance \!dimenA by #1
  \!dimenB=-\!plotyorigin \advance \!dimenB by #2
  \kern\!dimenA\raise\!dimenB\copy\!plotsymbol\kern-\!dimenA%
  \!countE=\!dimenA
  \!countF=\!dimenB
  \immediate\write\!replotfile{\the\!countE,\the\!countF.}%
  \ignorespaces}

\def\savelinesandcurves on "#1" {%
  \immediate\closeout\!replotfile
  \immediate\openout\!replotfile=#1%
  \let\!plot=\!!!plot}

\def\dontsavelinesandcurves {%
  \let\!plot=\!!plot}
\dontsavelinesandcurves

{\catcode`\%=11\xdef\!Commentsignal{
\def\writesavefile#1 {%
  \immediate\write\!replotfile{\!Commentsignal #1}%
  \ignorespaces}

\def\replot"#1" {%
  \expandafter\!replot\input #1 /}
\def\!replot#1,#2. {%
  \!dimenA=#1sp
  \kern\!dimenA\raise#2sp\copy\!plotsymbol\kern-\!dimenA
  \futurelet\!nextchar\!!replot}
\def\!!replot{%
  \if /\!nextchar
    \def\!next{\!finish}%
  \else
    \def\!next{\!replot}%
  \fi
  \!next}

\def\!Pythag#1#2#3{%
  \!dimenE=#1\relax
  \ifdim\!dimenE<\!zpt
    \!dimenE=-\!dimenE
  \fi
  \!dimenF=#2\relax
  \ifdim\!dimenF<\!zpt
    \!dimenF=-\!dimenF
  \fi
  \advance \!dimenF by \!dimenE
  \ifdim\!dimenF=\!zpt
    \!dimenG=\!zpt
  \else
    \!divide{8\!dimenE}\!dimenF\!dimenE
    \advance\!dimenE by -4pt
      \!dimenE=2\!dimenE
    \!removept\!dimenE\!!t
    \!dimenE=\!!t\!dimenE
    \advance\!dimenE by 64pt
    \divide \!dimenE by 2
    \!dimenH=7pt
    \!!Pythag\!!Pythag\!!Pythag
    \!removept\!dimenH\!!t
    \!dimenG=\!!t\!dimenF
    \divide\!dimenG by 8
  \fi
  #3=\!dimenG
  \ignorespaces}

\def\!!Pythag{
  \!divide\!dimenE\!dimenH\!dimenI
  \advance\!dimenH by \!dimenI
    \divide\!dimenH by 2}

\def\placehypotenuse for <#1> and <#2> in <#3> {%
  \!Pythag{#1}{#2}{#3}}

\def\!qjoin (#1,#2) (#3,#4){%
  \advance\!intervalno by 1
  \!ifcoordmode
    \edef\!xmidpt{#1}\edef\!ymidpt{#2}%
  \else
    \!dimenA=#1\relax \edef\!xmidpt{\the\!dimenA}%
    \!dimenA=#2\relax \edef\!ymidpt{\the\!dimenA}%
  \fi
  \!xM=\!M{#1}\!xunit  \!yM=\!M{#2}\!yunit   \!rotateaboutpivot\!xM\!yM
  \!xE=\!M{#3}\!xunit  \!yE=\!M{#4}\!yunit   \!rotateaboutpivot\!xE\!yE
  \!dimenA=\!xM  \advance \!dimenA by -\!xS
  \!dimenB=\!xE  \advance \!dimenB by -\!xM
  \!xB=3\!dimenA \advance \!xB by -\!dimenB
  \!xC=2\!dimenB \advance \!xC by -2\!dimenA
  \!dimenA=\!yM  \advance \!dimenA by -\!yS%
  \!dimenB=\!yE  \advance \!dimenB by -\!yM%
  \!yB=3\!dimenA \advance \!yB by -\!dimenB%
  \!yC=2\!dimenB \advance \!yC by -2\!dimenA%
  \!xprime=\!xB  \!yprime=\!yB
  \!dxprime=.5\!xC  \!dyprime=.5\!yC
  \!getf \!midarclength=\!dimenA
  \!getf \advance \!midarclength by 4\!dimenA
  \!getf \advance \!midarclength by \!dimenA
  \divide \!midarclength by 12
  \!arclength=\!dimenA
  \!getf \advance \!arclength by 4\!dimenA
  \!getf \advance \!arclength by \!dimenA
  \divide \!arclength by 12
  \advance \!arclength by \!midarclength
  \global\advance \totalarclength by \!arclength
  \ifdim\!distacross>\!arclength
    \advance \!distacross by -\!arclength
  \else
    \!initinverseinterp
    \loop\ifdim\!distacross<\!arclength
      \!inverseinterp
      \!xpos=\!t\!xC \advance\!xpos by \!xB
        \!xpos=\!t\!xpos \advance \!xpos by \!xS
      \!ypos=\!t\!yC \advance\!ypos by \!yB
        \!ypos=\!t\!ypos \advance \!ypos by \!yS
      \!plotifinbounds
      \advance\!distacross \plotsymbolspacing
      \!advancedashing
    \repeat
    \advance \!distacross by -\!arclength
  \fi
  \!xS=\!xE
  \!yS=\!yE
  \ignorespaces}

\def\!getf{\!Pythag\!xprime\!yprime\!dimenA%
  \advance\!xprime by \!dxprime
  \advance\!yprime by \!dyprime}

\def\!initinverseinterp{%
  \ifdim\!arclength>\!zpt
    \!divide{8\!midarclength}\!arclength\!dimenE
    \ifdim\!dimenE<\!wmin \!setinverselinear
    \else
      \ifdim\!dimenE>\!wmax \!setinverselinear
      \else
        \def\!inverseinterp{\!inversequad}\ignorespaces
         \!removept\!dimenE\!Ew
         \!dimenF=-\!Ew\!dimenE
         \advance\!dimenF by 32pt
         \!dimenG=8pt
         \advance\!dimenG by -\!dimenE
         \!dimenG=\!Ew\!dimenG
         \!divide\!dimenF\!dimenG\!beta
         \!gamma=1pt
         \advance \!gamma by -\!beta
      \fi
    \fi
  \fi
  \ignorespaces}

\def\!inversequad{%
  \!divide\!distacross\!arclength\!dimenG
  \!removept\!dimenG\!v
  \!dimenG=\!v\!gamma
  \advance\!dimenG by \!beta
  \!dimenG=\!v\!dimenG
  \!removept\!dimenG\!t}

\def\!setinverselinear{%
  \def\!inverseinterp{\!inverselinear}%
  \divide\!dimenE by 8 \!removept\!dimenE\!t
  \!countC=\!intervalno \multiply \!countC 2
  \!countB=\!countC     \advance \!countB -1
  \!countA=\!countB     \advance \!countA -1
  \wlog{\the\!countB th point (\!xmidpt,\!ymidpt) being plotted
    doesn't lie in the}%
  \wlog{ middle third of the arc between the \the\!countA th
    and \the\!countC th points:}%
  \wlog{ [arc length \the\!countA\space to \the\!countB]/[arc length
    \the \!countA\space to \the\!countC]=\!t.}%
  \ignorespaces}

\def\!inverselinear{%
  \!divide\!distacross\!arclength\!dimenG
  \!removept\!dimenG\!t}

\def\startrotation{%
  \let\!rotateaboutpivot=\!!rotateaboutpivot
  \let\!rotateonly=\!!rotateonly
  \!ifnextchar{b}{\!getsincos }%
    {\!getsincos by {\!cosrotationangle} {\!sinrotationangle} }}
\def\!getsincos by #1 #2 {%
  \edef\!cosrotationangle{#1}%
  \edef\!sinrotationangle{#2}%
  \!ifcoordmode
    \let\!ROnext=\!ccheckforpivot
  \else
    \let\!ROnext=\!dcheckforpivot
  \fi
  \!ROnext}
\def\!ccheckforpivot{%
  \!ifnextchar{a}{\!cgetpivot}%
    {\!cgetpivot about {\!xpivotcoord} {\!ypivotcoord} }}
\def\!cgetpivot about #1 #2 {%
  \edef\!xpivotcoord{#1}%
  \edef\!ypivotcoord{#2}%
  \!xpivot=#1\!xunit  \!ypivot=#2\!yunit
  \ignorespaces}
\def\!dcheckforpivot{%
  \!ifnextchar{a}{\!dgetpivot}{\ignorespaces}}
\def\!dgetpivot about #1 #2 {%
  \!xpivot=#1\relax  \!ypivot=#2\relax
  \ignorespaces}

\def\stoprotation{%
  \let\!rotateaboutpivot=\!!!rotateaboutpivot
  \let\!rotateonly=\!!!rotateonly
  \ignorespaces}

\def\!!rotateaboutpivot#1#2{%
  \!dimenA=#1\relax  \advance\!dimenA -\!xpivot
  \!dimenB=#2\relax  \advance\!dimenB -\!ypivot
  \!dimenC=\!cosrotationangle\!dimenA
    \advance \!dimenC -\!sinrotationangle\!dimenB
  \!dimenD=\!cosrotationangle\!dimenB
    \advance \!dimenD  \!sinrotationangle\!dimenA
  \advance\!dimenC \!xpivot  \advance\!dimenD \!ypivot
  #1=\!dimenC  #2=\!dimenD
  \ignorespaces}

\def\!!rotateonly#1#2{%
  \!dimenA=#1\relax  \!dimenB=#2\relax
  \!dimenC=\!cosrotationangle\!dimenA
    \advance \!dimenC -\!rotsign\!sinrotationangle\!dimenB
  \!dimenD=\!cosrotationangle\!dimenB
    \advance \!dimenD  \!rotsign\!sinrotationangle\!dimenA
  #1=\!dimenC  #2=\!dimenD
  \ignorespaces}
\def\!rotsign{}
\def\!!!rotateaboutpivot#1#2{\relax}
\def\!!!rotateonly#1#2{\relax}
\stoprotation

\def\!reverserotateonly#1#2{%
  \def\!rotsign{-}%
  \!rotateonly{#1}{#2}%
  \def\!rotsign{}%
  \ignorespaces}

\def\setshadegrid{%
  \!ifnextchar{s}{\!getspan }
    {\!getspan span <\!dshade>}}
\def\!getspan span <#1>{%
  \!dshade=#1\relax
  \!ifcoordmode
    \let\!GRnext=\!GRccheckforAP
  \else
    \let\!GRnext=\!GRdcheckforAP
  \fi
  \!GRnext}
\def\!GRccheckforAP{%
  \!ifnextchar{p}{\!cgetanchor }
    {\!cgetanchor point at {\!xshadesave} {\!yshadesave} }}
\def\!cgetanchor point at #1 #2 {%
  \edef\!xshadesave{#1}\edef\!yshadesave{#2}%
  \!xshade=\!xshadesave\!xunit  \!yshade=\!yshadesave\!yunit
  \ignorespaces}
\def\!GRdcheckforAP{%
  \!ifnextchar{p}{\!dgetanchor}%
    {\ignorespaces}}
\def\!dgetanchor point at #1 #2 {%
  \!xshade=#1\relax  \!yshade=#2\relax
  \ignorespaces}

\def\setshadesymbol{%
  \!ifnextchar<{\!setshadesymbol}{\!setshadesymbol<,,,> }}

\def\!setshadesymbol <#1,#2,#3,#4> (#5#6){%
  \!setputobject{#5}{#6}%
  \setbox\!shadesymbol=\box\!putobject%
  \!shadesymbolxshift=\!xshift \!shadesymbolyshift=\!yshift
  \!dimenA=\!xshift \advance\!dimenA \!smidge
  \!override\!dimenA{#1}\!lshrinkage%
  \!dimenA=\!wd \advance \!dimenA -\!xshift
    \advance\!dimenA \!smidge
    \!override\!dimenA{#2}\!rshrinkage
  \!dimenA=\!dp \advance \!dimenA \!yshift
    \advance\!dimenA \!smidge
    \!override\!dimenA{#3}\!bshrinkage
  \!dimenA=\!ht \advance \!dimenA -\!yshift
    \advance\!dimenA \!smidge
    \!override\!dimenA{#4}\!tshrinkage
  \ignorespaces}
\def\!smidge{-.2pt}%

\def\!override#1#2#3{%
  \edef\!!override{#2}%
  \ifx \!!override\empty
    #3=#1\relax
  \else
    \if z\!!override
      #3=\!zpt
    \else
      \ifx \!!override\!blankz
        #3=\!zpt
      \else
        #3=#2\relax
      \fi
    \fi
  \fi
  \ignorespaces}
\def\!blankz{ z}

\setshadesymbol ({\rm\vpt .})

\def\!startvshade#1(#2,#3,#4){%
  \let\!!xunit=\!xunit%
  \let\!!yunit=\!yunit%
  \let\!!xshade=\!xshade%
  \let\!!yshade=\!yshade%
  \def\!getshrinkages{\!vgetshrinkages}%
  \let\!setshadelocation=\!vsetshadelocation%
  \!xS=\!M{#2}\!!xunit
  \!ybS=\!M{#3}\!!yunit
  \!ytS=\!M{#4}\!!yunit
  \!shadexorigin=\!xorigin  \advance \!shadexorigin \!shadesymbolxshift
  \!shadeyorigin=\!yorigin  \advance \!shadeyorigin \!shadesymbolyshift
  \ignorespaces}

\def\!starthshade#1(#2,#3,#4){%
  \let\!!xunit=\!yunit%
  \let\!!yunit=\!xunit%
  \let\!!xshade=\!yshade%
  \let\!!yshade=\!xshade%
  \def\!getshrinkages{\!hgetshrinkages}%
  \let\!setshadelocation=\!hsetshadelocation%
  \!xS=\!M{#2}\!!xunit
  \!ybS=\!M{#3}\!!yunit
  \!ytS=\!M{#4}\!!yunit
  \!shadexorigin=\!xorigin  \advance \!shadexorigin \!shadesymbolxshift
  \!shadeyorigin=\!yorigin  \advance \!shadeyorigin \!shadesymbolyshift
  \ignorespaces}

\def\!lattice#1#2#3#4#5{%
  \!dimenA=#1
  \!dimenB=#2
  \!countB=\!dimenB
  \!dimenC=#3
  \advance\!dimenC -\!dimenA
  \!countA=\!dimenC
  \divide\!countA \!countB
  \ifdim\!dimenC>\!zpt
    \!dimenD=\!countA\!dimenB
    \ifdim\!dimenD<\!dimenC
      \advance\!countA 1 
    \fi
  \fi
  \!dimenC=\!countA\!dimenB
    \advance\!dimenC \!dimenA
  #4=\!countA
  #5=\!dimenC
  \ignorespaces}

\def\!qshade#1(#2,#3,#4)#5(#6,#7,#8){%
  \!xM=\!M{#2}\!!xunit
  \!ybM=\!M{#3}\!!yunit
  \!ytM=\!M{#4}\!!yunit
  \!xE=\!M{#6}\!!xunit
  \!ybE=\!M{#7}\!!yunit
  \!ytE=\!M{#8}\!!yunit
  \!getcoeffs\!xS\!ybS\!xM\!ybM\!xE\!ybE\!ybB\!ybC
  \!getcoeffs\!xS\!ytS\!xM\!ytM\!xE\!ytE\!ytB\!ytC
  \def\!getylimits{\!qgetylimits}%
  \!shade{#1}\ignorespaces}

\def\!lshade#1(#2,#3,#4){%
  \!xE=\!M{#2}\!!xunit
  \!ybE=\!M{#3}\!!yunit
  \!ytE=\!M{#4}\!!yunit
  \!dimenE=\!xE  \advance \!dimenE -\!xS
  \!dimenC=\!ytE \advance \!dimenC -\!ytS
  \!divide\!dimenC\!dimenE\!ytB
  \!dimenC=\!ybE \advance \!dimenC -\!ybS
  \!divide\!dimenC\!dimenE\!ybB
  \def\!getylimits{\!lgetylimits}%
  \!shade{#1}\ignorespaces}

\def\!getcoeffs#1#2#3#4#5#6#7#8{%
  \!dimenC=#4\advance \!dimenC -#2
  \!dimenE=#3\advance \!dimenE -#1
  \!divide\!dimenC\!dimenE\!dimenF
  \!dimenC=#6\advance \!dimenC -#4
  \!dimenH=#5\advance \!dimenH -#3
  \!divide\!dimenC\!dimenH\!dimenG
  \advance\!dimenG -\!dimenF
  \advance \!dimenH \!dimenE
  \!divide\!dimenG\!dimenH#8
  \!removept#8\!t
  #7=-\!t\!dimenE
  \advance #7\!dimenF
  \ignorespaces}

\def\!shade#1{%
  \!getshrinkages#1<,,,>\!nil
  \advance \!dimenE \!xS
  \!lattice\!!xshade\!dshade\!dimenE
    \!parity\!xpos
  \!dimenF=-\!dimenF
    \advance\!dimenF \!xE
  \!loop\!not{\ifdim\!xpos>\!dimenF}
    \!shadecolumn%
    \advance\!xpos \!dshade
    \advance\!parity 1
  \repeat
  \!xS=\!xE
  \!ybS=\!ybE
  \!ytS=\!ytE
  \ignorespaces}

\def\!vgetshrinkages#1<#2,#3,#4,#5>#6\!nil{%
  \!override\!lshrinkage{#2}\!dimenE
  \!override\!rshrinkage{#3}\!dimenF
  \!override\!bshrinkage{#4}\!dimenG
  \!override\!tshrinkage{#5}\!dimenH
  \ignorespaces}
\def\!hgetshrinkages#1<#2,#3,#4,#5>#6\!nil{%
  \!override\!lshrinkage{#2}\!dimenG
  \!override\!rshrinkage{#3}\!dimenH
  \!override\!bshrinkage{#4}\!dimenE
  \!override\!tshrinkage{#5}\!dimenF
  \ignorespaces}

\def\!shadecolumn{%
  \!dxpos=\!xpos
  \advance\!dxpos -\!xS
  \!removept\!dxpos\!dx
  \!getylimits
  \advance\!ytpos -\!dimenH
  \advance\!ybpos \!dimenG
  \!yloc=\!!yshade
  \ifodd\!parity
     \advance\!yloc \!dshade
  \fi
  \!lattice\!yloc{2\!dshade}\!ybpos%
    \!countA\!ypos
  \!dimenA=-\!shadexorigin \advance \!dimenA \!xpos
  \loop\!not{\ifdim\!ypos>\!ytpos}
    \!setshadelocation
    \!rotateaboutpivot\!xloc\!yloc%
    \!dimenA=-\!shadexorigin \advance \!dimenA \!xloc
    \!dimenB=-\!shadeyorigin \advance \!dimenB \!yloc
    \kern\!dimenA \raise\!dimenB\copy\!shadesymbol \kern-\!dimenA
    \advance\!ypos 2\!dshade
  \repeat
  \ignorespaces}

\def\!qgetylimits{%
  \!dimenA=\!dx\!ytC
  \advance\!dimenA \!ytB
  \!ytpos=\!dx\!dimenA
  \advance\!ytpos \!ytS
  \!dimenA=\!dx\!ybC
  \advance\!dimenA \!ybB
  \!ybpos=\!dx\!dimenA
  \advance\!ybpos \!ybS}

\def\!lgetylimits{%
  \!ytpos=\!dx\!ytB
  \advance\!ytpos \!ytS
  \!ybpos=\!dx\!ybB
  \advance\!ybpos \!ybS}

\def\!vsetshadelocation{
  \!xloc=\!xpos
  \!yloc=\!ypos}
\def\!hsetshadelocation{
  \!xloc=\!ypos
  \!yloc=\!xpos}

\def\!axisticks {%
  \def\!nextkeyword##1 {%
    \expandafter\ifx\csname !ticks##1\endcsname \relax
      \def\!next{\!fixkeyword{##1}}%
    \else
      \def\!next{\csname !ticks##1\endcsname}%
    \fi
    \!next}%
  \!axissetup
    \def\!axissetup{\relax}%
  \edef\!ticksinoutsign{\!ticksinoutSign}%
  \!ticklength=\longticklength
  \!tickwidth=\linethickness
  \!gridlinestatus
  \!setticktransform
  \!maketick
  \!tickcase=0
  \def\!LTlist{}%
  \!nextkeyword}

\def\ticksout{%
  \def\!ticksinoutSign{+}}

\ticksout

\def\nogridlines{%
  \def\!gridlinestatus{\!gridlinestoofalse}}
\nogridlines

\def\loggedticks{%
  \def\!setticktransform{\let\!ticktransform=\!logten}}
\def\unloggedticks{%
  \def\!setticktransform{\let\!ticktransform=\!donothing}}
\def\!donothing#1#2{\def#2{#1}}
\unloggedticks

\expandafter\def\csname !ticks/\endcsname{%
  \!not {\ifx \!LTlist\empty}
    \!placetickvalues
  \fi
  \def\!tickvalueslist{}%
  \def\!LTlist{}%
  \expandafter\csname !axis/\endcsname}

\def\!maketick{%
  \setbox\!boxA=\hbox{%
    \beginpicture
      \!setdimenmode
      \setcoordinatesystem point at {\!zpt} {\!zpt}
      \linethickness=\!tickwidth
      \ifdim\!ticklength>\!zpt
        \putrule from {\!zpt} {\!zpt} to
          {\!ticksinoutsign\!tickxsign\!ticklength}
          {\!ticksinoutsign\!tickysign\!ticklength}
      \fi
      \if!gridlinestoo
        \putrule from {\!zpt} {\!zpt} to
          {-\!tickxsign\!xaxislength} {-\!tickysign\!yaxislength}
      \fi
    \endpicturesave <\!Xsave,\!Ysave>}%
    \wd\!boxA=\!zpt}

\def\!ticksin{%
  \def\!ticksinoutsign{-}%
  \!maketick
  \!nextkeyword}

\def\!ticksout{%
  \def\!ticksinoutsign{+}%
  \!maketick
  \!nextkeyword}

\def\!tickslength<#1> {%
  \!ticklength=#1\relax
  \!maketick
  \!nextkeyword}

\def\!tickslong{%
  \!tickslength<\longticklength> }

\def\!ticksshort{%
  \!tickslength<\shortticklength> }

\def\!tickswidth<#1> {%
  \!tickwidth=#1\relax
  \!maketick
  \!nextkeyword}

\def\!ticksandacross{%
  \!gridlinestootrue
  \!maketick
  \!nextkeyword}

\def\!ticksbutnotacross{%
  \!gridlinestoofalse
  \!maketick
  \!nextkeyword}

\def\!tickslogged{%
  \let\!ticktransform=\!logten
  \!nextkeyword}

\def\!ticksunlogged{%
  \let\!ticktransform=\!donothing
  \!nextkeyword}

\def\!ticksunlabeled{%
  \!tickcase=0
  \!nextkeyword}

\def\!ticksnumbered{%
  \!tickcase=1
  \!nextkeyword}

\def\!tickswithvalues#1/ {%
  \edef\!tickvalueslist{#1! /}%
  \!tickcase=2
  \!nextkeyword}

\def\!ticksquantity#1 {%
  \ifnum #1>1
    \!updatetickoffset
    \!countA=#1\relax
    \advance \!countA -1
    \!ticklocationincr=\!axisLength
      \divide \!ticklocationincr \!countA
    \!ticklocation=\!axisstart
    \loop \!not{\ifdim \!ticklocation>\!axisend}
      \!placetick\!ticklocation
      \ifcase\!tickcase
          \relax 
        \or
          \relax 
        \or
          \expandafter\!gettickvaluefrom\!tickvalueslist
          \edef\!tickfield{{\the\!ticklocation}{\!value}}%
          \expandafter\!listaddon\expandafter{\!tickfield}\!LTlist%
      \fi
      \advance \!ticklocation \!ticklocationincr
    \repeat
  \fi
  \!nextkeyword}

\def\!ticksat#1 {%
  \!updatetickoffset
  \edef\!Loc{#1}%
  \if /\!Loc
    \def\next{\!nextkeyword}%
  \else
    \!ticksincommon
    \def\next{\!ticksat}%
  \fi
  \next}

\def\!ticksfrom#1 to #2 by #3 {%
  \!updatetickoffset
  \edef\!arg{#3}%
  \expandafter\!separate\!arg\!nil
  \!scalefactor=1
  \expandafter\!countfigures\!arg/
  \edef\!arg{#1}%
  \!scaleup\!arg by\!scalefactor to\!countE
  \edef\!arg{#2}%
  \!scaleup\!arg by\!scalefactor to\!countF
  \edef\!arg{#3}%
  \!scaleup\!arg by\!scalefactor to\!countG
  \loop \!not{\ifnum\!countE>\!countF}
    \ifnum\!scalefactor=1
      \edef\!Loc{\the\!countE}%
    \else
      \!scaledown\!countE by\!scalefactor to\!Loc
    \fi
    \!ticksincommon
    \advance \!countE \!countG
  \repeat
  \!nextkeyword}

\def\!updatetickoffset{%
  \!dimenA=\!ticksinoutsign\!ticklength
  \ifdim \!dimenA>\!offset
    \!offset=\!dimenA
  \fi}

\def\!placetick#1{%
  \if!xswitch
    \!xpos=#1\relax
    \!ypos=\!axisylevel
  \else
    \!xpos=\!axisxlevel
    \!ypos=#1\relax
  \fi
  \advance\!xpos \!Xsave
  \advance\!ypos \!Ysave
  \kern\!xpos\raise\!ypos\copy\!boxA\kern-\!xpos
  \ignorespaces}

\def\!gettickvaluefrom#1 #2 /{%
  \edef\!value{#1}%
  \edef\!tickvalueslist{#2 /}%
  \ifx \!tickvalueslist\!endtickvaluelist
    \!tickcase=0
  \fi}
\def\!endtickvaluelist{! /}

\def\!ticksincommon{%
  \!ticktransform\!Loc\!t
  \!ticklocation=\!t\!!unit
  \advance\!ticklocation -\!!origin
  \!placetick\!ticklocation
  \ifcase\!tickcase
    \relax 
  \or 
    \ifdim\!ticklocation<-\!!origin
      \edef\!Loc{$\!Loc$}%
    \fi
    \edef\!tickfield{{\the\!ticklocation}{\!Loc}}%
    \expandafter\!listaddon\expandafter{\!tickfield}\!LTlist%
  \or 
    \expandafter\!gettickvaluefrom\!tickvalueslist
    \edef\!tickfield{{\the\!ticklocation}{\!value}}%
    \expandafter\!listaddon\expandafter{\!tickfield}\!LTlist%
  \fi}

\def\!separate#1\!nil{%
  \!ifnextchar{-}{\!!separate}{\!!!separate}#1\!nil}
\def\!!separate-#1\!nil{%
  \def\!sign{-}%
  \!!!!separate#1..\!nil}
\def\!!!separate#1\!nil{%
  \def\!sign{+}%
  \!!!!separate#1..\!nil}
\def\!!!!separate#1.#2.#3\!nil{%
  \def\!arg{#1}%
  \ifx\!arg\!empty
    \!countA=0
  \else
    \!countA=\!arg
  \fi
  \def\!arg{#2}%
  \ifx\!arg\!empty
    \!countB=0
  \else
    \!countB=\!arg
  \fi}

\def\!countfigures#1{%
  \if #1/%
    \def\!next{\ignorespaces}%
  \else
    \multiply\!scalefactor 10
    \def\!next{\!countfigures}%
  \fi
  \!next}

\def\!scaleup#1by#2to#3{%
  \expandafter\!separate#1\!nil
  \multiply\!countA #2\relax
  \advance\!countA \!countB
  \if -\!sign
    \!countA=-\!countA
  \fi
  #3=\!countA
  \ignorespaces}

\def\!scaledown#1by#2to#3{%
  \!countA=#1\relax
  \ifnum \!countA<0 
    \def\!sign{-}
    \!countA=-\!countA
  \else
    \def\!sign{}%
  \fi
  \!countB=\!countA
  \divide\!countB #2\relax
  \!countC=\!countB
    \multiply\!countC #2\relax
  \advance \!countA -\!countC
  \edef#3{\!sign\the\!countB.}
  \!countC=\!countA 
  \ifnum\!countC=0 
    \!countC=1
  \fi
  \multiply\!countC 10
  \!loop \ifnum #2>\!countC
    \edef#3{#3\!zero}%
    \multiply\!countC 10
  \repeat
  \edef#3{#3\the\!countA}
  \ignorespaces}

\def\!placetickvalues{%
  \advance\!offset \tickstovaluesleading
  \if!xswitch
    \setbox\!boxA=\hbox{%
      \def\\##1##2{%
        \!dimenput {##2} [B] (##1,\!axisylevel)}%
      \beginpicture
        \!LTlist
      \endpicturesave <\!Xsave,\!Ysave>}%
    \!dimenA=\!axisylevel
      \advance\!dimenA -\!Ysave
      \advance\!dimenA \!tickysign\!offset
      \if -\!tickysign
        \advance\!dimenA -\ht\!boxA
      \else
        \advance\!dimenA  \dp\!boxA
      \fi
    \advance\!offset \ht\!boxA
      \advance\!offset \dp\!boxA
    \!dimenput {\box\!boxA} [Bl] <\!Xsave,\!Ysave> (\!zpt,\!dimenA)
  \else
    \setbox\!boxA=\hbox{%
      \def\\##1##2{%
        \!dimenput {##2} [r] (\!axisxlevel,##1)}%
      \beginpicture
        \!LTlist
      \endpicturesave <\!Xsave,\!Ysave>}%
    \!dimenA=\!axisxlevel
      \advance\!dimenA -\!Xsave
      \advance\!dimenA \!tickxsign\!offset
      \if -\!tickxsign
        \advance\!dimenA -\wd\!boxA
      \fi
    \advance\!offset \wd\!boxA
    \!dimenput {\box\!boxA} [Bl] <\!Xsave,\!Ysave> (\!dimenA,\!zpt)
  \fi}

\normalgraphs
\catcode`!=12 

\catcode`@=11 \catcode`!=11

\let\!pictexendpicture=\endpicture
\let\!pictexframe=\frame
\let\!pictexlinethickness=\linethickness
\let\!pictexmultiput=\multiput
\let\!pictexput=\put

\def\beginpicture{%
  \setbox\!picbox=\hbox\bgroup%
  \let\endpicture=\!pictexendpicture
  \let\frame=\!pictexframe
  \let\linethickness=\!pictexlinethickness
  \let\multiput=\!pictexmultiput
  \let\put=\!pictexput
  \let\input=\@@input   
  \!xleft=\maxdimen
  \!xright=-\maxdimen
  \!ybot=\maxdimen
  \!ytop=-\maxdimen}

\let\frame=\!latexframe

\let\pictexframe=\!pictexframe

\let\linethickness=\!latexlinethickness
\let\pictexlinethickness=\!pictexlinethickness

\let\\=\@normalcr
\catcode`@=12 \catcode`!=12

\language0 
\textheight=21.5cm \topmargin=0.8cm \footskip=1.5cm
\oddsidemargin1.1cm \evensidemargin1.1cm
\newif\iffinalplot \finalplottrue 
\newcommand{\h}{\hspace{0.09em}}
\newcommand{\hh}{\hspace{0.05em}}
\newcommand{\bnabla}{{\mbox{\boldmath $\nabla$}}}
\begin{document}
\title{The two-fluid model with superfluid entropy}
\author{{}\\R. Schaefer and T. Fliessbach \\
University of Siegen,
Fachbereich Physik\\ D-57068 Siegen, Germany}
\date{}
\maketitle
\vspace*{2.5cm}
\thispagestyle{empty}

$$\mbox{{\bf Abstract}}$$
The two-fluid model of liquid helium is generalized to the case that
the superfluid fraction has a small entropy content. We present
theoretical arguments in favour of such a small superfluid entropy. In
the generalized two-fluid model various sound modes of He$\;$II are
investigated. In a superleak carrying a persistent current the
superfluid entropy leads to a new sound mode which we call sixth sound.
The relation between the sixth sound and the superfluid entropy is
discussed in detail.

\smallskip

\vspace*{2cm}

\noindent PACS numbers: 67.40.Bz, 67.40.Pm, 67.40.Kh

\vfill

\noindent Published in Nuovo Cimento {\bf 16\,D} (1994) 373

\newpage

\section{Introduction}

The two-fluid model can be considered as the fundamental theory for the
hydro\-dynamics of liquid helium below the $\lambda$-point. One of the
model assumptions is that the entropy of the superfluid fraction
vanishes. A small superfluid entropy $S_{\rm s}$ (of the order of one
percent of the total entropy $S$) is, however, not excluded by the
experiment. In this paper we generalize the two-fluid model to the case
of a small superfluid entropy and investigate the consequences.

The paper {\em Entropy of the Superfluid Component of Helium}\/
by Glick and Werntz \cite{gl69} finds that the ratio $S_{\rm s}/S$ is
less than $3\%$; earlier experiments cited in Ref. \cite{gl69} are less
accurate. A newer experiment which is sensitive to this point has been
performed by Singsaas and Ahlers \cite{si84}. In this work, second
sound measurements are interpreted as that of entropy. As a result, no
difference to the true (caloric) entropy has been found. Due to the
uncertainty in the absolute values of the caloric entropy (about 1 to
2\%) the limit for $S_{\rm s}/S$ is not much lower than that of Ref.
\cite{gl69}.

In nearly all theoretical approaches $S_{\rm s}$ is taken to be zero.
As discussed by Putterman \cite{pu74} this is an {\em assumption}\/.
Microscopically, $S_{\rm s}=0$ follows from the identification of the
superfluid density $\rho_{\rm s}$ with the square of a macroscopic wave
function \cite{lo54}. This identification is plausible but unproven.
In section 1.2 we present theoretical arguments in favour of a small
superfluid entropy.

The remainder of this paper is based on the hypothesis of a
non-vanishing superfluid entropy of a size which is not excluded by the
experiment. We generalize the two-fluid model for this case (section
2). In the generalized two-fluid model we determine various sound modes
(section 3). We find a new sound mode (sixth sound) which exists only
if the superfluid entropy does not vanish. Section 4 discusses in
detail an experiment by which the sixth sound could be detected. If
the 6th sound exists this experiment would determine the superfluid
entropy $S_{\rm s}$. If the 6th sound does not exist this experiment
would yield a considerably lower upper limit for $S_{\rm s}$.

\subsection{Two-fluid model}
We start with a short review of the two-fluid model. Based on ideas by
Tisza the two-fluid model was developed by Landau \cite{la41}. We use
Putterman's monograph \cite{pu74} as a standard reference for this
model. Without dissipative terms the two-fluid equations read
\begin{equation}
\partial_t\h \rho + \bnabla (\rho_{\rm n}\h {\bf u}_{\rm n}+ \rho_{\rm
s}\h {\bf u}_{\rm s} ) = 0\,,
\end{equation}
\begin{equation}
\partial_t \h (\rho \h s) + \bnabla (\rho\h s\h {\bf
u}_{\rm n}) = 0\,,
\end{equation}
\begin{equation}
\partial_t\h ( \rho_{\rm n}\h {\bf u}_{\rm n} + \rho_{\rm
s}\h {\bf u}_{\rm s} )_i + \partial_j \h\big(
P\h\delta_{ij} + \rho_{\rm n}\h u_{{\rm n} i} \h u_{{\rm n} j}
        + \rho_{\rm s}\h u_{{\rm s} i} \h u_{{\rm s} j}
\big) = 0 \,,
\end{equation}
\begin{equation}
m\, \partial_t\h {\bf u}_{\rm s} +
m\h ({\bf u}_{\rm s} \bnabla )\h {\bf u}_{\rm s} = - \bnabla \mu \,.
\end{equation}
We use the abbreviations $\partial_t=\partial /\partial t$ and
$\partial_i = \partial /\partial x_i$. In eq. (3) we sum over the index
$j$. The entropy per particle is denoted by $s=S/N$, the mass of a helium
atom by $m$, the mass density by $\rho = m\h N/V$, the velocity field
by ${\bf u}$, the pressure by $P$ and the chemical potential by $\mu$.
The indices n and s refer to the normal and superfluid phase,
respectively.

Eq. (1) represents the mass conservation, eq. (2) the entropy
conservation (for reversible processes), eq. (3) the momentum
conservation or Euler equation, and eq. (4) the motion of the
superfluid. The eight equations (1) -- (4) describe the dynamics of
eight independent macroscopic variables. As independent variables one
might chose the temperature $T$, the pressure $P$ and the velocities
${\bf u}_{\rm n}$ and ${\bf u}_{\rm s}$. These variables are fields
depending on the coordinate ${\bf r}$ and on the time $t$.

All macroscopic quantities $X$ are functions of the eight variables.
Galilean invariant macroscopic quantities can be written as a function
of three variables only,
\begin{equation}
X = X (T,P,w^2) \quad \mbox{where~ }
{\bf w} = {\bf u}_{\rm n} - {\bf u}_{\rm s} \,.
\end{equation}
Most results (like the sound velocities) are eventually expressed by
equilibrium quantities $X(T,P,0)$ depending on $T$ and $P$ only.

The eqs. (1) -- (4) are supplemented by dissipative terms and by
Onsager's quantization rule. The resulting two-fluid model is the
fundamental theory for the hydrodynamics of He$\;$II. We restrict
ourselves mainly to eqs. (1) -- (4) for which we discuss the
modification due to $S_{\rm s}\ne 0$.

The two-fluid model is a phenomenological macroscopic theory
\cite{pu74}. It is based on macroscopic conservation laws (like mass
conservation and the first and second law of thermodynamics) and on
experimental evidence (like $S_{\rm s}\approx 0$). There are, however,
also theoretical ideas on a microscopic level which support these
equations. We refer in particular to London's postulate \cite{lo54,ti90}
that the superfluid consists of a macroscopic number of particles
moving coherently in a single quantum state. This conception explains
that the superfluid entropy vanishes which is implicitly assumed in
(2). Furthermore, it implies that the superfluid velocity is a
gradient field, or that
\begin{equation}
{\rm curl}\,{\bf u}_{\rm s} = 0 \,.
\end{equation}
This statement is contained in (4). The condition (6) may be violated
in a vortex. For a vortex line London's macroscopic wave function leads
to Onsager's quantization rule.

\subsection{Superfluid entropy}

In this subsection we present arguments in favour of a small superfluid
entropy $S_{\rm s}$. These arguments lead to a theoretical estimate of
$S_{\rm s}$.

The non-existence of the $\lambda$-transition in ${}^3$He proves that
this transition is due to the exchange symmetry of the ${}^4$He-atoms.
The interatomic forces are about the same in liquid ${}^3$He and
${}^4$He; they are not the cause of the $\lambda$-transition. These
facts suggest a close connection between the Bose-Einstein-condensation
of the ideal Bose gas (IBG) and the $\lambda$-transition of liquid
$^4$He. This point of view has been put forward by London
\cite{lo38} and has subsequently been supported by several authors
\cite{fe53,pu74}.

The condensate of the IBG forms a macroscopic wave function and
provides thus a model for the superfluid motion. At this point there
is, however, a serious discrepancy between liquid helium and the IBG.
The critical behaviour of the condensate fraction of the IBG is
\begin{equation}
\frac{\rho_0}{\rho} \propto |t|^{\hh 2\beta},\qquad
\beta = \frac{1}{2} \;,
\end{equation}
where $t = (T - T_\lambda) / T_\lambda$ is the relative temperature. In
contrast to (7) the superfluid fraction of He$\,$II$\h$ behaves roughly
like
\begin{equation}
\frac{\rho_{\rm s}}{\rho} \propto |t|^{\hh 2\hh\nu},\qquad
\nu\approx \frac{1}{3} \;.
\end{equation}
Just below the transition this implies $\rho_0\ll \rho_{\rm
s}$. The critical exponent $\beta =1/2$ is an essential feature of the
IBG. It does not appear possible to change this value by some
modification of the IBG. (Note that the IBG free energy is not a
logical starting point for a renormalization procedure because it
already applies to an infinite system.)

We present now a possible scheme which reconciles (7) with (8)
preserving at the same time the essential features of the IBG (like
$\beta =1/2$).

Following Chester \cite{ch55} we use the IBG wave function together
with a Jastrow factor $F =\Pi f_{ij}$; such an ansatz is based on
Feynman's discussion \cite{fe53}. Allowing for a condensate motion the
many-body wave function reads
\begin{equation}
\Psi = {\cal S}\h F\h \big[ \exp(\h {\rm i}\h \Phi ) \big]^{n_0}
\prod_{ {\bf k}\h \ne \h 0} \left[\, \varphi_{\bf k} \,\right]^{n_k} .
\end{equation}
Here ${\cal S}$ denotes the symmetrization operator. The $\varphi_{\bf
k}$ are the real single particle functions of the non-condensed
particles and the $n_k$ are the occupation numbers. The schematic
notation $[\varphi_{\bf k} ]^{n_k}$ stands for the product
$\varphi_{\bf k}({\bf r}_1) \cdot\varphi_{\bf k}({\bf r}_2)\cdot
\ldots\cdot \varphi_{\bf k}({\bf r}_{n_k}) $; this notation applies
also to $[ \exp(\h {\rm i}\h \Phi )]^{n_0}$. All $n_0$ particles adopt
the same phase factor $ \exp(\h {\rm i}\h \Phi ({\bf r}))$ forming the
macroscopic wave funtion
\begin{equation}
\psi ({\bf r}) =\sqrt{ \frac{n_0}{V}\,}\,
 \exp(\h {\rm i}\h \Phi ({\bf r}))\,.
\end{equation}
This implies that the condensate particles move coherently with the
(small) velocity ${\bf u}_{\rm s} = \hbar\, \bnabla \Phi /m$.

Eqs. (9) and (10) are a standard description \cite{lo54} for a
superfluid motion in an IBG-like model. (Actually, one has to construct
a suitable coherent state \cite{an66} instead of (9). This point is,
however, not essential for the following discussion.) In this
description the superfluid fraction $\rho_{\rm s}/ \rho$ equals the
condensate fraction $n_0/N = \rho_0 / \rho$. We note that the current
$\rho_0\h {\bf u}_{\rm s}$ is not depleted by the real Jastrow factors
(in contrast to the condensate density $\rho_0$ itself).

In order to dissolve the discrepancy between (7) and (8) we assume that
{\em non-condensed particles move coherently with the condensate}\/.
This is possible if non-condensed particles adopt the macroscopic phase
of the condensate:
\begin{equation}
\Psi = {\cal S}\h F\h \big[ \exp(\h {\rm i}\h \Phi )\big]^{n_0}
\prod_{k\h \le \h k_{\rm c}} \left[\,\varphi_{\bf k}\,\exp(\h {\rm i}\h
\Phi ) \,\right]^{n_k} \prod_{k\h > \h k_{\rm c}} \left[\,
\varphi_{\bf k} \,\right]^{n_k} .
\end{equation}
For the single particle states with $n_k\gg 1$ this phase ordering
requires only a very small entropy change. The ansatz (11) assumes
therefore this phase ordering for low momentum states. The limit
$k_{\rm c}$ up to which the particles move coherently may be considered
as a model parameter.

We evaluate the quantum mechanical expectation value $\langle\Psi|\h
\widehat{j}\h |\Psi \rangle $ of the current operator $\widehat{j}$ for
(11). The result is proportional to ${\bf u}_{\rm s} = \hbar\, \bnabla
\Phi /m$; neither the real Jastrow factor nor the real single particle
functions contribute. Equating the statistical expectation value of
this current with $\rho_{\rm s}
\h {\bf u}_{\rm s}$ yields
\begin{equation}
\frac{\rho_{\rm s}}{\rho} = \frac{\rho_0}{\rho} +
\frac{1}{N}\sum_{k\h \le\h k_{\rm c}} \langle n_k\rangle\,.
\end{equation}
For $\langle n_k\rangle$ we use the occupation numbers of the IBG-form.
Fitting (12) to the experimental superfluid density determines $k_{\rm
c}(t)$. In this way the discrepancy between (7) and (8) is removed.
The asymptotic behaviour $\rho_{\rm s}\propto |t|^{2/3}$
implies $k_{\rm c}\propto |t|^{2/3}$.

With (12) fitted to the experimental superfluid density, the entropy of
the contibuting non-condensed particles can be calculated. The
resulting prediction \cite{fl91} of this superfluid entropy $S_{\rm s}$
is shown in Fig. 1. Due to $k_{\rm c}\propto |t|^{2/3}$ the superfluid
entropy per particle, $s_{\rm s} = S_{\rm s}/N_{\rm s} \propto k_{\rm
c}^{\,2}\propto |t|^{4/3}$, vanishes for $T\to T_\lambda$,
\begin{equation}
s_{\rm s}(T_\lambda ,P) =0 \,.
\end{equation}
Because of $\rho_0/\rho \to 1$ it vanishes also for $T\to 0$.

\begin{figure}[t]
\begin{center} \setlength{\unitlength}{10mm}
\begin{picture}(13.7,6)(-.2,-0.5) \thicklines
\iffinalplot
\small
\put(-.13,0.01){\beginpicture
\setquadratic
\setcoordinatesystem units <5.4cm,2.7cm>
\setplotarea x from 0 to 2.4, y from 0 to 2.4
\axis bottom ticks in withvalues 0 0.05 0.1 0.15 0.2 /
at 0 0.5 1 1.5 2 / /
\axis left ticks in numbered from 0.5 to 2 by 0.5 /
\axis right ticks in from 0.5 to 2 by 0.5 /
\axis top ticks in withvalues 2.1 2.0 1.9 1.8 1.7 /
at 0.3315 0.7919 1.2538 1.7157 2.1776 / /
\setplotsymbol ({\viiipt\rm .})
 \plot 
  0.000  0.000   0.010  0.077   0.020  0.161   0.030  0.243
  0.040  0.322   0.050  0.398   0.060  0.471   0.070  0.540
  0.080  0.607   0.090  0.670   0.100  0.730   0.200  1.203
  0.300  1.494   0.400  1.658   0.500  1.732   0.600  1.739
  0.700  1.699   0.800  1.623   0.900  1.523   1.000  1.407
  1.100  1.280   1.200  1.149   1.300  1.016   1.400  0.885
  1.500  0.760   1.600  0.641   1.700  0.530   1.800  0.429
  1.900  0.338   2.000  0.257   2.100  0.188   2.200  0.129
  2.300  0.080 /
\setdashpattern <2mm,1.33mm> 
 \plot
  0.000  0.000   0.010  0.002   0.020  0.006   0.030  0.012
  0.040  0.019   0.050  0.028   0.060  0.037   0.070  0.047
  0.080  0.058   0.090  0.069   0.100  0.081   0.200  0.212
  0.300  0.343   0.400  0.460   0.500  0.554   0.600  0.625
  0.700  0.672   0.800  0.698   0.900  0.704   1.000  0.693
  1.100  0.668   1.200  0.631   1.300  0.585   1.400  0.532
  1.500  0.475   1.600  0.415   1.700  0.355   1.800  0.297
  1.900  0.240   2.000  0.188   2.100  0.141   2.200  0.099
  2.300  0.063 /
\endpicture }
\large
\put(3.2,5){$s_{\rm s}/s$}
\put(4.7,2.17){$S_{\rm s}/S$}
\normalsize
\put(12.4,-0.55){$|t|$}
\put(-0.05,6){{\%}}
\put(10.55,6.8){$T({\rm K})$}
\fi
\end{picture} \end{center}
{\bf Figure 1}: Prediction \cite{fl91} of the superfluid density
$S_{\rm s}(T,P_{\rm SVP})$ as a function of the temperature and at
saturated vapour pressure.
\end{figure}

The experimental superfluid fraction $\rho_{\rm s}/\rho$ obeys rather
well the law of corresponding states \cite{ma76}; this means that it
can be written as a function of $t = t(T,P) = T/T_\lambda (P) -1$
alone. Since $s_{\rm s}$ is determined by a fit to $\rho_{\rm s}/\rho$
this should also hold for the superfluid entropy,
\begin{equation}
s_{\rm s}(T,P) \approx g(t) = g(T/T_\lambda (P) -1)\,.
\end{equation}
In this way the full $T$- and $P$-dependence follows from the
prediction shown in Fig. 1 (using the lower scale only). The superfluid
entropy (14) is of the form (5) with $w=0$.

The presented argument for a small superfluid entropy is based on the
obvious relation between the Bose-Einstein-condensation and the
$\lambda$-transition, together with the discrepancy between (7) and
(8). This argument serves as a motivation of our investigation.  The
remainder of this paper could just as well be based on the mere
hypothesis of a non-vanishing superfluid entropy.

\section{Modified two-fluid model}

In this section we generalize the two-fluid model to the case of a
small superfluid entropy $s_{\rm s} = s_{\rm s}(T,P,w^2)$.

\subsection{Mass, momentum and entropy conservation}
In eqs. (1) and (3) the normal and the superfluid part are treated in a
symmetric way. Therefore these equations are unchanged by a
non-vanishing superfluid entropy:
\begin{equation}
\partial_t\h \rho + \bnabla (\rho_{\rm n}\h {\bf u}_{\rm n}+ \rho_{\rm
s}\h {\bf u}_{\rm s} ) = 0 \quad\mbox{for $s_{\rm s}\ne 0$}\,,
\end{equation}
\begin{equation}
\partial_t\h ( \rho_{\rm n}\h {\bf u}_{\rm n} + \rho_{\rm
s}\h {\bf u}_{\rm s} )_i + \partial_j \h\big(
P\h\delta_{ij} + \rho_{\rm n}\h u_{{\rm n} i} \h u_{{\rm n} j}
        + \rho_{\rm s}\h u_{{\rm s} i} \h u_{{\rm s} j}
\big) = 0 \quad\mbox{for $s_{\rm s}\ne 0$} \,.
\end{equation}
In eq. (2) the entropy density $\rho\h s$ is carried by the normal
fraction alone; consequently, the entropy current density is $\rho\h s\,
{\bf u}_{\rm n}$. For $s_{\rm s}\ne 0$ the entropy density is partly
carried by the superfluid fraction leading to a different entropy
current density:
\begin{equation}
\rho\h s\h {\bf u}_{\rm n} \; \longrightarrow\;
\rho_{\rm n}\h s_{\rm n}\h {\bf u}_{\rm n} +
\rho_{\rm s}\h s_{\rm s}\h {\bf u}_{\rm s}\,.
\end{equation}
Here $s_{\rm s} = S_{\rm s} /N_{\rm s}$ and $s_{\rm n} = (S- S_{\rm
s}) /(N - N_{\rm s})$. The entropy continuity equation reads now
\begin{equation}
\partial_t \h (\rho \h s) + \bnabla\hh (\rho_{\rm n}\h s_{\rm n} \h
{\bf u}_{\rm n} + \rho_{\rm s}\h s_{\rm s}\h {\bf u}_{\rm s}) = 0
\quad\mbox{for $s_{\rm s}\ne 0$}\,.
\end{equation}

\subsection{Equation of superfluid motion}

In the static limit eq. (4) becomes $\bnabla \mu = -s\, \bnabla T + v\,
\bnabla P = 0$ where $v = V/N = m/\rho$. This yields the well-known
fountain pressure (FP)
\begin{equation}
\left(\frac{dP}{dT}\right)_{\!\rm FP} = \frac{s}{v}
\quad\mbox{ for $s_{\rm s}= 0$}\,.
\end{equation}
The origin of the FP may be explained as follows: Consider two
containers with liquid He$\;$II connected by a superleak; initially
both containers have the same temperature $T$ and pressure $P$. By
increasing the pressure in one container some superfluid liquid is
pushed through the superleak to the other container. Since the
superfluid fraction carries no entropy the other container becomes
colder; effectively $dP>0$ in the first container is accompanied by
$dT>0$. According to (19) the ratio $dP/dT$ is proportional to the {\em
missing}\/ entropy $s$ per transferred particle. A possible superfluid
entropy $s_{\rm s} = S_{\rm s}/N_{\rm s} $ diminishes the missing
entropy per transferred particle, which means
\begin{equation}
s \; \longrightarrow \; s - s_{\rm s} \qquad
\begin{array}{l}
\mbox{(entropy deficit of a superfluid}\\
\mbox{~particle relative to average).}
\end{array}
\end{equation}
As in (17), this replacement reflects the change in the entropy
transport due to $s_{\rm s}\ne 0$. The replacement (20) applies,
however, to the static limit and does not immediately yield the wanted
generalization of (4).

A basic property of the superfluid motion is ${\rm curl}\, {\bf u}_{\rm
s} = 0$, eq. (6). This property is well-established experimentally.
Theoretically, it follows from the conception that the supervelocity is
the gradient of a macroscopic phase; this conception is not altered by
the modified picture presented in section 1.2. Eq. (6) implies that the
l.h.s. of (4) is a gradient field. Therefore, the generalization of (4)
must be of the form
\begin{equation}
m\, \partial_t \h {\bf u}_{\rm s} +
m\h ({\bf u}_{\rm s} \bnabla )\h {\bf u}_{\rm s} =
 - \bnabla (\mu - \mu_{\rm s}) \quad\mbox{for $s_{\rm s}\ne 0$}\,.
\end{equation}
The chemical potential $\mu$ yields the contribution $ - \partial
\mu/\partial T = s $ in (19). Consequently, the replacement (20)
implies
\begin{equation}
\frac{\partial \mu_{\rm s} (T,P,w^2)}{\partial T} = - s_{\rm
s}(T,P,w^2) \,.
\end{equation}
Together with $s_{\rm s}$, eq. (13), the discussed modifications should
vanish at $T_\lambda$. Therefore
\begin{equation}
\mu _{\rm s}(T,P,w^2) = - \int_{T_\lambda}^T dT'\; s_{\rm s}(T',P,w^2)
\,.
\end{equation}
Eq. (21) with (23) defines the generalization of (4).

We determine the FP from the generalized equation of motion (21). The
FP experiment is done for ${\bf u}_{\rm n} ={\bf u}_{\rm s} =0$ or
$w^2=0$. Then eq. (21) yields $\bnabla (\mu -\mu_{\rm s})=0$ and
\begin{equation}
\left(\frac{dP}{dT}\right)_{\!\rm FP} =
\, \frac{s-s_{\rm s}}{v-v_{\rm s}}\,
\approx \, \frac{s-s_{\rm s}}{v}
\end{equation}
where $s_{\rm s} = s_{\rm s} (T,P,\h 0)$ and
\begin{equation}
v_{\rm s} = \frac{\partial \mu_{\rm s} (T,P,\h 0)}{\partial P} =
- \int_{T_\lambda}^T dT'\; \frac{\partial s_{\rm s} (T',P,\h 0)}
{\partial P}\,.
\end{equation}
The quantity $v_{\rm s}$ is rather small and may be neglected in (24).
{}From (14) we obtain $T_\lambda\, (\partial s_{\rm s}/ \partial P) = -
\h T \, (\partial s_{\rm s}/ \partial T )\, dT_\lambda /dP$. Using
this, $dT_\lambda /dP\approx -0.01\,{\rm K/bar}$ and the entropy
$s_{\rm s}$ of Fig. 1, the integral (25) can be evaluated numerically.
The resulting $|v_{\rm s}/v|$ has a similar temperature dependence as
$s_{\rm s}/s$. The absolute values of $|v_{\rm s}/v|$ are much smaller,
\begin{equation}
\left| \frac{v_{\rm s}}{v} \right|\, \le \, 3\cdot 10^{-4} .
\end{equation}
The term $v_{\rm s}$ has not been considered in previous
generalizations \cite{di49,fl91} of (19). It is derived theoretically;
it follows from ${\rm curl}\, {\bf u}_{\rm s} = 0$ and the pressure
dependence of the superfluid entropy.

\subsection{Summary}
The two-fluid equations with $s_{\rm s}\ne 0$ are
\begin{equation}
\partial_t\h \rho + \bnabla (\rho_{\rm n}\h {\bf u}_{\rm n}+ \rho_{\rm
s}\h {\bf u}_{\rm s} ) = 0\,,
\end{equation}
\begin{equation}
\partial_t \h (\rho \h s) + \bnabla (\rho_{\rm n}\h s_{\rm n}\h {\bf
u}_{\rm n} + \rho_{\rm s}\h s_{\rm s}\h {\bf u}_{\rm s} ) = 0\,,
\end{equation}
\begin{equation}
\partial_t\h ( \rho_{\rm n}\h {\bf u}_{\rm n} + \rho_{\rm
s}\h {\bf u}_{\rm s} )_i + \partial_j \h\big(
P\h\delta_{ij} + \rho_{\rm n}\h u_{{\rm n} i} \h u_{{\rm n} j}
        + \rho_{\rm s}\h u_{{\rm s} i} \h u_{{\rm s} j}
\big) = 0 \,,
\end{equation}
\begin{equation}
m\, \partial_t\h {\bf u}_{\rm s} +
m\h ({\bf u}_{\rm s} \bnabla )\h {\bf u}_{\rm s} = - \bnabla (\mu
-\mu_{\rm s} )
\end{equation}
where
\begin{equation}
\mu _{\rm s}(T,P,w^2) = - \int_{T_\lambda}^T dT'\; s_{\rm s}(T',P,w^2)
\,.
\end{equation}
For $s_s =0$ these equations reduce to (1) -- (4). They are still eight
equations for eight variables. The superfluid entropy $s_{\rm s}$ is
just a further macroscopic quantity; as any other Galilean invariant
macroscopic quantity it is of the form (5). It does not constitute a
new independent variable.

In addition we note:
\begin{enumerate}
\item
The equations (27) -- (30) have to be supplemented by terms describing
dissipative effects. These terms might also contain corrections of the
size ${\cal O}(s_{\rm s}/s)$ relative to their well-known form
\cite{pu74}. For practical purposes (like an estimate of the damping of
sound modes) we will use unmodified dissipative terms.
\item
Eq. (6) follows from (30). Onsager's quantization rule is unchanged.
\end{enumerate}
Together with the two-fluid equations the underlying microscopic
conception is slightly modified (section 1.2). As usual there is a
macroscopic wave function $\psi = \sqrt{\rho_0 \,}\h \exp(\h {\rm i}\h
\Phi ({\bf r}))$ which determines the supervelocity ${\bf u}_{\rm s} =
\hbar\, \bnabla \Phi /m$. What is given up is the identification of
$|\psi|^2$ with $\rho_{\rm s}$. At the same time London's main point,
namely the relation between the IBG and liquid helium, is reinforced by
reconciling (7) with (8).

\section{Sound modes}

\subsection{Introduction}

A major and important application of hydrodynamic equations is the
evaluation of sound modes. This application is of particular interest
for testing the modified two-fluid model because sound velocities can
be measured with high accuracy. We consider the first and second sound
of bulk He$\;$II (section 3.2) and the fourth sound of clamped He$\;$II
(section 3.3). We determine the corrections in the sound velocities due
to a non-vanishing superfluid entropy. The derivation of the fourth
sound for $s_{\rm s}\ne 0$ leads nearly automatically to a new sound
mode which we call sixth sound. The detailed calculations are given in
Ref. \cite{sc93}. The sixth sound has already been presented in a short
letter \cite{sc94}.

The standard ansatz for sound modes
\begin{eqnarray}
T({\bf r},t) & = & T_0\; + \;\Delta T\;
\exp(\h{\rm i}\h({\bf k}\h{\bf r} -\omega\h t))\,,
\\
P({\bf r},t) & = & P_0\; + \;\Delta P\,
\exp(\h{\rm i}\h({\bf k}\h{\bf r} -\omega\h t))\,,
\\
{\bf u}_{\rm n}({\bf r},t) & = & {\bf u}_{{\rm n},0} + \Delta {\bf
u}_{\rm n}\h \exp(\h{\rm i}\h({\bf k}\h{\bf r} -\omega\h
t)) \,,
\\
{\bf u}_{\rm s}({\bf r},t) & = & {\bf u}_{{\rm s},0}\h + \hh \Delta
{\bf u}_{\rm s}\,\exp(\h{\rm i}\h({\bf k}\h{\bf r} -\omega\h t))
\end{eqnarray}
is inserted into (27) -- (30). The constant values $T_0$, $P_0$, ${\bf
u}_{{\rm n},0}$ and ${\bf u}_{{\rm s},0}$ solve the equations.
Quadratic and higher order terms in the (small) amplitudes $\Delta T$,
$\Delta P$, $\Delta {\bf u}_{\rm n}$ and $\Delta {\bf u}_{\rm s}$ are
omitted. This yields a linear, homogeneous system of equations for the
amplitudes. For a non-trivial solution the determinant of the
coefficient matrix must vanish. This condition yields solutions of the
form $\omega_\nu = \omega_\nu (k)$ (for a specific direction of ${\bf
k}$). The ratio
\begin{equation}
c_\nu =  \frac{\omega_\nu (k)}{k}
\end{equation}
is the velocity of the sound wave. The sound velocities of the common
two-fluid model (1) -- (4) are denoted by $c_{\nu ,0}$. Because of
\begin{equation}
\frac{s_{\rm s}}{s}\le 2\cdot 10^{-2}\,, \qquad
\left|\frac{v_{\rm s}}{v}\right|\le 3\cdot 10^{-4}
\end{equation}
the differences between $c_\nu$ and $c_{\nu ,0}$ are expected to be
small.

All coefficients in the linearized equations are taken at $T_0$, $P_0$
and $w_0^{\,2} = ({\bf u}_{{\rm n},0}-{\bf u}_{{\rm s},0})^2 $. The
final results are expressed by thermodynamic quantities at $w_0 = 0$.
Eventually we omit the index zero and use the notation
\begin{equation}
X(T,P) = X(T_0,P_0,\h 0)\,.
\end{equation}

\subsection{First and second sound}

The equilibrium state of bulk He$\;$II has a given temperature $T_0$,
pressure $P_0$ and
\begin{equation}
{\bf u}_{{\rm n},0} = {\bf u}_{{\rm s},0} = 0\,.
\end{equation}
We insert (32) -- (35) into (27) -- (30) and determine (36). This
calculation is quite analogous to that \cite{pu74,ti90} in the common
two-fluid model. Therefore, we restrict ourselves to a presentation of
the results. Because of (39) the coefficients in the linearized
equations and the sound velocities are of the form (38).

For the first and second sound the leading corrections to $c_{1,0}$ and
$c_{2,0}$ are given by
\begin{equation}
c_1(T,P) = c_{1,0} \; \left(
\,1 - \frac{s_{\rm s}}{s} \, \left( 1- \frac{c_V}{c_P} \right)
\frac{u_2^{\;2}}{u_1^{\;2}}
- \frac{ v_{\rm s} }{v}\;
\frac{ T \left( s_{\rm n} - s\right)} { 2\h \rho \, c_P} \;
\left(\frac{\partial \rho}{\partial T}\right)_{\! P}\, \right)
\end{equation}
and
\begin{equation}
c_2(T,P) = c_{2,0} \left( 1- \frac{s_{\rm s}}{s} \right)\,.
\end{equation}
We use the abbreviations
\begin{equation}
u_1 = \sqrt{ \frac{\partial P (\rho ,s)}{\partial \rho} }
\quad \mbox{and}\quad
u_2 = \sqrt{ \frac{T\h \rho_{\rm s}\, s^2}{m\, \rho_{\rm n} \, c_V}}\;.
\end{equation}
By $c_V$ and $c_P$ we denote the specific heats at constant volume and
pressure, respectively. The standard approximations for $c_{1,0}$ and
$c_{2,0}$ are
\begin{equation}
c_{1,0} \approx u_1\quad \mbox{and} \quad c_{2,0} \approx u_2 \;
\sqrt{\frac{c_V}{c_P}}\;.
\end{equation}
The corrections to these approximations are discussed in detail in Ref.
\cite{pu74}; they are of the relative order $(u_2^{\,2}/u_1^{\,2})\h
(1- c_V/c_P)\sim 10^{-5}$.

There are two correction terms in (40). The first one has the relative
size $2\cdot 10^{-7}$, the second one $2\cdot 10^{-8}$. These
corrections are of academic interest only; they are even smaller than
the error in $c_{1,0} \approx u_1$.

The corrections for the second sound are of the order $s_{\rm s}/s \le
2\%$; they might be observable. Using (24) we may define a `fountain
pressure' entropy $s_{\mbox{\tiny FP}}$ by
\begin{equation}
\frac{s_{\mbox{\tiny FP}}}{v} = \left(\frac{dP}{dT}\right)_{\!\rm FP}\,.
\end{equation}
This yields a common expression for $c_2$ and $c_{2,0}$:
\begin{equation}
 \sqrt{ \frac {T\h \rho_{\rm s}\h s_{\mbox{\tiny FP}}^{\,2} } {m\,
\rho_{\rm n} \, c_P}} \, = \left\{ \begin{array}{cll} c_{2,0} &&
(s_{\rm s} =0)
\\[3mm] c_{2} && (s_{\rm s} \ne 0)\,. \end{array}\right.
\end{equation}

\subsection{Fourth and sixth sound}

In this subsection we consider He$\;$II in which the normal phase is
clamped,
\begin{equation}
{\bf u}_{\rm n}({\bf r},t) = 0\,.
\end{equation}
This reduces the number of variables to five, for which we may choose
$T$, $P$ and ${\bf u}_{\rm s}$. Their dynamics is determined by the
five equations (27), (28) and (30); eq. (29) is effectively replaced by
(46).

A standard device for measuring sound modes in clamped helium is a ring
filled with powder (Fig. 2). The position (middle line) of the ring may
be described by
\begin{equation}
{\bf r}_{\rm ring} = ( R\h \cos\phi ,\, R\h \sin\phi ,\, 0)\,,
\qquad \phi = 0,\ldots , \hh  2\pi\,.
\end{equation}
The thickness of the ring is assumed to be small compared to the radius
$R$. Then the ${\bf r}$-dependence of the considered fields reduces to
a ${\phi}$-dependence and the supervelocity is parallel to the ring,
\begin{equation}
{\bf u}_{\rm s}({\bf r},t) = u_{\rm s}(\phi,t)\; {\bf e}_\phi .
\end{equation}
This reduces the number of variables and equations from five to three.
Using (46) and (48) the field equations (27), (28) and (30) become
\begin{eqnarray}
\partial_t \h \rho + \frac{1}{R} \; \frac{\partial}{\partial\phi}
\left( \rho_{\rm s} \h u_{\rm s} \right) & = & 0 \,,
\\
\partial_t \h ( \rho\h s) +
\frac{1}{R} \; \frac{\partial}{\partial\phi}
\;(\rho_{\rm s}\h s_{\rm s}\h u_{\rm s}) & = & 0 \,,
\\
m\,\partial_t\h u_{\rm s} + \frac{1}{R}\;\frac{\partial}{\partial\phi}
\left( \mu-\mu_{\rm s} + \frac{m \h u_{\rm s} ^2}{2}\; \right) &=& 0 \,.
\end{eqnarray}

\begin{figure}[t]
\begin{center} \unitlength=1cm
\begin{picture}(6,5.3)(-3,-3.4) \thicklines
\iffinalplot
\put(-3.8,2.96){Powder-filled ring with He$\,$II}
\thinlines
\put(0,0){\circle*{0.13}}
\put(0,0){\line(1,0){0.85}}
\put(1.4,0){\line(1,0){0.85}}
\put(0.95,-.1){$R$}
\put(2.25,-0.1){\line(0,1){0.2}}
\put(-1.9,2.8){\vector(1,-1){0.85}}
\put(-.12,0){
\beginpicture
\setcoordinatesystem units <\unitlength,\unitlength>
\setplotsymbol ({\xpt\rm .})
\circulararc 360 degrees from 0 -1.98 center at 0 0
\circulararc 360 degrees from 0 -2.5 center at 0 0
\circulararc -10 degrees from 0 -3 center at 0 0
\circulararc 20 degrees from 0 -3 center at 0 0
\arrow <3mm> [0.2,0.4] from 1 -2.82 to 1.3 -2.67
\setshadegrid span <0.5mm>
\setquadratic
\vshade
 -2.51 0.000 0.000 <,z,,>  -2.250 -1.090 1.090  -2 -1.5 1.5 /
\vshade
     -2.000 0.000 1.500
 <z,z,,> -1.750 0.968 1.785  -1.500 1.323 2.000
 <z,z,,> -1.000 1.732 2.291  -0.500 1.936 2.449
 <z,z,,> 0.000 2.000 2.500  0.500 1.936 2.449
 <z,z,,> 1.000 1.732 2.291  1.500 1.323 2.000
 <z,z,,> 1.750 0.968 1.785  2.000 0.000 1.500 /

\vshade
     -2.000 -1.500 0.000
 <z,z,,> -1.750 -1.785 -0.968  -1.500 -2.000 -1.323
 <z,z,,> -1.000 -2.291 -1.732  -0.500 -2.449 -1.936
 <z,z,,>  0.000 -2.500 -2.000  0.500 -2.449 -1.936
 <z,z,,>  1.000 -2.291 -1.732  1.500 -2.000 -1.323
 <z,z,,>  1.750 -1.785 -0.968  2.000 -1.500 0.000 /
\vshade
  2.000 -1.5 1.5 <z,,,> 2.250 -1.09 1.09  2.500 0.000 0.000 /
\endpicture }
\footnotesize
\put(-0.75,-2.83){persistent}
\put(-0.75,-3.28){current}
\normalsize
\put(1.45,-2.65){${\bf u}_{\rm s}$}
\small
\put(-4,0.2){$T+\Delta T$}
\put(-4,-0.3){$P+\Delta P$}
\put(2.75,0.2){$T$}
\put(2.75,-0.3){$P$}
\fi
\end{picture} \end{center}
{\bf Figure 2}: A ring with clamped helium is a standard device for fourth
sound and persistent current experiments. We consider the solutions of
the linearized equations of motion for this configuration. For $s_{\rm
s}= 0$ these solutions are the fourth sound and a static fountain
pressure (FP) gradient $\Delta P/\Delta T $. For $s_{\rm s}\ne 0$ (and
in the presence of a persistent current) the FP gradient becomes a
sound mode which we call sixth sound.
\end{figure}

\noindent The following derivation is simplified by using the
variables
\begin{equation}
T,\quad \mu_{\rm c} =  \mu-\mu_{\rm s} + \frac{m \h u_{\rm s} ^2}{2}
 \quad \mbox{and} \quad j = \rho_{\rm s}\h u_{\rm s}
\end{equation}
rather than $T$, $P$ and $u_{\rm s}$. Taking into account (46), the
specific geometry, and the new variables, the ansatz for sound modes
reads
\begin{eqnarray}
T(\phi ,t) & = & T_0\, + \, \Delta T \,
\exp (\h {\rm i}\h [\, k R\h \phi - \omega\h t ] )\,,
\\
\mu_{\rm c}(\phi ,t) & = & \! \mu_{{\rm c}, 0} + \Delta \mu_{\rm
c} \exp (\h {\rm i}\h [\, k R\h \phi - \omega\h t ] )\,,
\\
j(\phi ,t) & = & \, j_0 \,\h + \,\hh \Delta j \,
\,\exp (\h {\rm i}\h [\, k R\h \phi - \omega\h t ] )\,.
\end{eqnarray}
The equilibrium state has the constant values $T_0$, $\mu_{{\rm c},0} $
and
\begin{equation}
j_0 =  \rho_{{\rm s},0} \h  u_{{\rm s},0} = \mbox{const.} \,.
\end{equation}
A persistent current ($j_0 \ne 0$) is a metastable equilibrium.

Because of $T(\phi ,t) = T(\phi+2\pi ,t)$ the possible $k$-values are
restricted to
\begin{equation}
k = \frac{n}{R} \quad \mbox{where}\quad
n \in \{ \pm 1 , \pm 2 , \pm 3 , \ldots \}.
\end{equation}
Inserting (53) -- (55) into (49) -- (51), the resulting linearized
equations are:
\small 
\begin{equation}
\!\!\left(\!\! \begin{array}{ccc}
\omega \; {\displaystyle\frac{\partial \rho}{\partial T}}
&
\omega \; {\displaystyle\frac{\partial \rho}{\partial \mu_{\rm c}}}
&
\omega\; {\displaystyle\frac{\partial \rho}{\partial j}} - k
\\[5mm]
\omega\h\rho\, {\displaystyle\frac{\partial s}{\partial T}}
- k\h j\, {\displaystyle\frac{\partial s_{\rm s}}{\partial T}}
&
\;\,
\omega \h \rho\, {\displaystyle\frac{\partial s}{\partial \mu_{\rm c}}}
- k\h j\,{\displaystyle\frac{\partial s_{\rm s}}{\partial \mu_{\rm
c}}}
\;\,
&
\omega \h \rho \, {\displaystyle\frac{\partial s}{\partial j}} +
 (s\! - \!s_{\rm s})\hh k - k\h j\,
 {\displaystyle\frac{\partial s_{\rm s}}{\partial j}}
\\[5mm]
\omega\, u_{\rm s}\;
{\displaystyle\frac{\partial\rho_{\rm s}}{\partial T}}
&
\omega \, u_{\rm s}\;
{\displaystyle\frac{\partial\rho_{\rm s}}{\partial \mu_{\rm s}}}
+ \rho_{\rm s}\, k
&
\omega \, u_{\rm s} \;
{\displaystyle\frac{\partial \rho_{\rm s}}{\partial j}}\, - \omega
\end{array}\!\! \right)\!
\left(\! \begin{array}{c}
\Delta T
\\[7mm]
\Delta \mu_{\rm c}
\\[7mm]
\Delta j
\end{array}
\raisebox{16mm}{\hh}\raisebox{-15.5mm}{\hh}\! \! \right) = 0\,.
\raisebox{-17mm}{\,}
\end{equation}
\normalsize
The first line represents eq. (49), the third line eq. (51). For the
second line we subtracted eq. (49) times $s$ from (50). All
coefficients in (58) are to be taken at $T_0$, $\mu_{{\rm c},0}$ and
$j_0$. Here and in the following we suppress the index zero.  In each
partial derivative two of the three variables $T$, $\mu_{\rm c}$ and
$j$ are kept constant.

For a non-trivial solution the determinant of the coefficient matrix in
(58) must vanish. This condition yields three eigenvalues $\omega_\nu$
(for given $k$), or three sound velocities $c_\nu$. The three sound
velocities may be distinguished by their dependence on the velocity
$u_{{\rm s}}$:
\begin{equation}
c_\nu = \left\{ \begin{array}{ccl}
\pm \h c_4 &\propto & \, 1 +{\cal O}(u_{{\rm s}})
\\[2mm]
 c_6 &\propto & u_{{\rm s}} \left( 1 +{\cal O}(u_{{\rm s}}^{\;2})
\right) \,.
\end{array}\right.
\end{equation}
The solutions $\pm \h c_4$ are those of the well-known fourth sound. For
$s_{\rm s} \ne 0$ and $u_{\rm s} \ne 0$ we obtain a new sound mode
which we call {\em sixth sound}\/.

In the following we restrict ourselves to the leading order in $u_{{\rm
s}}$. Higher order contributions are necessarily small because
$|u_{{\rm s}}|$ must be less than the critical velocity $u_{\rm crit}$.
In a specific experiment, these contributions can be further suppressed
by choosing $|u_{{\rm s}}|\ll u_{\rm crit}$. The coefficients appearing
in the linearized equations are written as
\begin{equation}
X(T,\h \mu_{{\rm c}},\h j^2) = X(T,\h \mu_{{\rm c}}) + {\cal O}
(u_{{\rm s}}^{\,2})\,.
\end{equation}
The condition for a non-trivial solution of (58) yields
\begin{equation}
c_4 = \sqrt{ \, \frac{\rho_{\rm s}}{m\h \rho}
\;\frac{ \mbox{\small $\displaystyle
( s-s_{\rm s} )\; \frac{\partial \rho}{\partial T}
+ \rho\; \frac{\partial s}{\partial T} $} \raisebox{-3.8mm}{\,}}
{\mbox{\small $\displaystyle
\frac{\partial \rho}{\partial \mu_{\rm c}}\;
\frac{\partial s}{\partial T}
- \frac{\partial \rho}{\partial T}
\; \frac{\partial s}{\partial \mu_{\rm c}} \,
 $} \raisebox{6.4mm}{\,}} }
\end{equation}
and
\begin{equation}
c_6 = u_{\rm s}\;\frac{\rho_{\rm s} }{\rho}\;
\frac{ \mbox{\small $
\displaystyle{ \frac{\partial s_{\rm s}}{\partial T}}$}
\raisebox{-3.8mm}{\,}}
{\mbox{\small $ \displaystyle \frac{\partial s}{\partial T} +
 \frac{s-s_{\rm s}}{\rho} \; \frac{\partial \rho}{\partial T}$}
\raisebox{6.4mm}{\,}} \;.
\end{equation}
Since we restrict ourselves to the leading order in $u_{{\rm s}}$ all
quantities are to be taken at $T$, $\mu_{\rm c}$ and $j=0$, eq.  (60).
We switch now to the arguments $T$, $P$ and $w^2=0$ by writing
\begin{eqnarray}
\frac{\partial X}{\partial T} & = &
\frac{\partial X (T,\mu_{\rm c})} {\partial T} \,
= \, \left( \frac{\partial X}{\partial T} \right)_{\! P}
+ \frac{s-s_{\rm s}}{v-v_{\rm s}}\,
\left( \frac{\partial X}{\partial P} \right)_{\! T}\;,
\\[2mm]
\frac{\partial X }{\partial \mu_{\rm c}}
 & = &
\frac{\partial X(T,\mu_{\rm c})}{\partial \mu_{\rm c}}
\, = \, \frac{1}{v-v_{\rm s}} \;
\left( \frac{\partial X}{\partial P} \right)_{\! T} \;.
\end{eqnarray}
We use $(\partial \mu_{\rm c}/\partial T )_P = -( s - s_{\rm s})$ and
$(\partial \mu_{\rm c}/\partial P )_T = v - v_{\rm s}$ which follow
from (22) and (25). With (63) and (64) we obtain from (61) and (62):
\begin{equation}
c_4 =
 \sqrt{ \,
\frac{\rho_{\rm s}}{\rho} \left( 1-\frac{v_{\rm s}}{v} \right)
\left[ 1 + \frac{ T\h ( s - s_{\rm s}) }{\rho \,c_P }
\, \frac{2 - v_{\rm s} / v}{1 - v_{\rm s}/v}
\left( \frac{\partial \rho}{\partial T} \right)_{\! P}\,
\right] \, u_1^{\;2} +
\frac{\rho_{\rm n}}{\rho} \left( 1-\frac{s_{\rm s}}{s} \right)^{\! 2}
u_2^{\;2} }
\end{equation}
and
\begin{equation}
c_6 = u_{\rm s}\;\frac{\rho_{\rm s} }{\rho}\;
\frac{\mbox {\small $ \displaystyle
\left( \frac{\partial s_{\rm s}}{\partial T} \right)_{\! P}
+ \frac{s - s_{\rm s}}{v-v_{\rm s}}
\left( \frac{\partial s_{\rm s}}{\partial P} \right)_{\! T}
$ } \raisebox{-4.5mm}{\,}}
{\mbox{\small $ \displaystyle \left( \frac{\partial s}{\partial T}
\right)_{\! P} \left[
1+\frac{\rho_{\rm n}\hh u_2^{\;2} \left( 1-s_{\rm s} / s \right)^2}
    {\rho_{\rm s}\hh u_1^{\;2} \left( 1-v_{\rm s} / v \right) }
\right]
+ \frac{s-s_{\rm s}}{\rho}\,
\frac{2-v_{\rm s} / v }{1-v_{\rm s} / v }
\left( \frac{\partial \rho}{\partial T} \right)_{\! P}
$} } \;.
\end{equation}
These results are exact to any order in $s_{\rm s}/s$ and $v_{\rm
s}/v$, and to leading order in $u_{\rm s}$. We present now approximate
but simpler formulae. Neglecting the higher order corrections eq. (65)
becomes
\begin{equation}
c_4(T,P) =
 \sqrt{ \,  \frac{\rho_{\rm s}}{\rho} \left( 1-\frac{v_{\rm s}}{v} \right)
\left[ 1 + \frac{2\,  T \h (s-s_{\rm s})  }{\rho \,c_P }
\left( \frac{\partial \rho}{\partial T} \right)_{\! P}\,
\right] \, u_1^{\;2} +
\frac{\rho_{\rm n}}{\rho} \left( 1-\frac{s_{\rm s}}{s} \right)^{\! 2}
u_2^{\;2} }\,.\;
\end{equation}
For $v_{\rm s}=0$ and $s_{\rm s}=0$ this result is well-known
\cite{pu74}. The first term under the square root is the dominant one;
therefore  it is appropriate to retain the small correction $v_{\rm
s}/v$ in this term (as compared to the larger correction $s_{\rm s}/s$
in the second term). The correction in the second term corresponds to
that in (41); the one in the first term is different from that in (40).

The sound velocity $c_6$ is proportional to $s_{\rm s}$. For a more
handy expression we may therefore neglect all corrections to $c_6$ of
the order of one percent. We go back to (62) and use $\partial s_{\rm
s}/\partial T = ( \partial s_{\rm s}/\partial T )_{\mu_{\rm c}}\approx
(\partial s_{\rm s}/\partial T)_\mu$ and $\partial s/\partial T \approx
(\partial s /\partial T)_\mu$. Because
\begin{equation}
\left| \, \frac{s}{\rho}\;\frac{\partial \rho }{\partial T}\,\right|
\sim 10^{-2}\, \left|\frac {\partial s}{\partial T}\right| \,,
\end{equation}
the second term in the denominator in (62) can be omitted. Thus we
obtain
\begin{equation}
c_6(T,P) = u_{\rm s}\; \frac{\rho_{\rm s}}{ \rho}\;
\frac{c_{\mu ,{\rm s}}}{ c_\mu}
\end{equation}
where
\begin{eqnarray}
c_{\mu } &=& T\,\left( \frac{\partial s}{\partial T }\right)_{\!\!\mu}
\quad \mbox{and} \quad
c_{\mu ,{\rm s}} = T\,
\left( \frac{ \partial s_{\rm s} }{\partial T }\right)_{\!\!\mu}
\end{eqnarray}
are the specific heats at constant chemical potential. For the
practical evaluation (next section) one may use the specific heats at
constant pressure instead.

\subsection{Discussion of the sixth sound}
The fourth sound is experimentally and theoretically well-established.
The following discussion centers therefore on the sixth sound. We
derive the amplitudes of this sound mode and its damping.

For $s_{\rm s}=0$  the frequency of sixth sound becomes zero; the
solution (32), (33) represents a static temperature and pressure
gradient. For the discussion of this limit we may go back to the
equations of motion (49) -- (51): For $\partial_t =0$ the first
equation is solved by $\partial (\rho_{\rm s}\h u_{\rm s})/\partial
\phi = 0$, and the second one is trivially fulfilled. The third one
yields $\mu -\mu_{\rm s} =\mbox{const.} +{\cal O}(u_s^{\,2})$ or $\mu
\approx \mbox{const.}$. That means that the sixth sound reduces to a
static fountain pressure (FP) gradient in the limit $s_{\rm s}=0$. This
situation is changed for $s_{\rm s}\ne 0$: The second term in (50) is
now non-zero (provided that $u_{\rm s}\ne 0$) and leads via the first
term to a time-dependence. The originally static fountain pressure
gradient becomes an oscillating mode; effectively, the FP gradient is
shifted with velocity $c_6$ along the ring. The sixth sound may be
conceived as an entropy transport along the ring caused by the
persistent current (section 4).

As discussed, the sixth sound amplitudes solve eq. (51) by $\mu
\approx \mbox{const.}$. This means
\begin{equation}
\left(\frac{\Delta T}{\Delta P}\right)_{\!\rm 6th~sound}\; \approx
\;\frac{v}{s} \;.
\end{equation}
The corresponding ratio for the fourth sound is well-known (for $s_{\rm
s}= 0$). Neglecting corrections of the order $10^{-2}$ this ratio is
\begin{equation}
\left(\frac{\Delta T}{\Delta P}\right)_{\!\rm 4th~sound}\; \approx \,
\left[ \, \frac{\rho_{\rm n} \, u_2^{\;2}}{\rho_{\rm s}\, u_1^{\;2}} +
\frac{T \h s}{\rho \, c_P}
\left( \frac{\partial \rho}{\partial T} \right)_{\! P}
\, \right] \;\frac{v}{s} \, \ll \,\frac{v}{s} \;.
\end{equation}
The fourth and sixth sound can be distinguished by their amplitudes
and by their velocities: The temperature amplitude of the fourth sound
is much smaller than that of the sixth sound (for a given pressure
amplitude). The velocity of the sixth sound has a characteristic
proportionality to $ u_{{\rm s}}$.

In order to calculate the damping of the sound modes the two-fluid
equations are supplemented by (unmodified) dissipative terms
\cite{pu74}. The detailed calculation \cite{sc93} shows that only the
heat conductivity $\kappa $ contributes to the damping of the sixth
sound. We derive this damping in a simplified way. Taking into account
the heat conduction the first coefficient in the second row of (58)
becomes
\begin{equation}
\omega\, \rho \; \frac{\partial s}{\partial T}
- k\, j \; \frac{\partial s_{\rm s}}{\partial T}
+ {\rm i} \; \frac{m\h \kappa}{T}\; k^2 \approx 0 \,.
\end{equation}
For the sixth sound with its dominant $\Delta T$-amplitude in (58) this
coefficient must be approximately zero. For $\kappa =0$ this condition
yields the velocity $c_6 = \omega /k$ of (69). For $\kappa \ne 0$ we
obtain instead
\begin{equation}
\omega = c_6\h k -
\,{\rm i} \;\frac{m\h k^2\h \kappa }{\rho \,c_\mu }= \omega_{\rm FP} -
\,{\rm i}\, \Gamma_{\rm FP} \,.
\end{equation}
This result is of the same accuracy as (69). The real quantities $
\omega_{\rm FP} $ and $\Gamma_{\rm FP}$ will be used and discussed in
section 4.

\subsection{Detectability of a superfluid entropy}

Each of the calculated sound velocities has some corrections due to
$s_{\rm s}\ne 0$. We discuss whether a possible superfluid entropy may
be detected by measuring the sound velocity:
\begin{enumerate}

\item[]
{\bf First sound:} The relative difference between $c_1$ and $c_{1,0}$
is of the order $10^{-7}$. This is many orders below the experimental
accuracy (and also below the theoretical accuracy of $c_{1,0}$).

\item[]
{\bf Second sound:} The velocity $c_2$ has been measured with a
systematic error of less than $0.4\%$ \cite{si84}. Assuming the
validity of (19) this measurement has been interpreted as that of the
entropy. This entropy may be called `fountain pressure' entropy $s_{\rm
FP}$, eqs. (44) and (45). According to Fig. 1 we expect a 1 to 2\%{}
difference between $s_{\rm FP}$ and the true (caloric) entropy $s$.
Unfortunately the absolute value of $s$ at the $\lambda$-point is also
uncertain by about 2\%{}.

Improving the present experimental accuracy by some factor (say 5) one
should be able to see a difference between $s$ and $s_{\rm FP}$, in
particular because of the steep rise of $s_{\rm s}$ just below the
$\lambda$-point (Fig. 1). This possibility has been discussed in more
detail in Ref. \cite{fl91}.

\item[]
{\bf Fourth sound:} The fourth sound velocity (67) is dominated by the
first term under the square root. The correction in this term is of
the order $10^{-4}$; it is smaller than the experimental accuracy.

\item[]
{\bf Sixth sound:} The sixth sound velocity is proportional to the
superfluid entropy. The sixth sound is therefore the prime candidate
for detecting and measuring the superfluid entropy. Its observability
will be discussed in detail in section 4.

\end{enumerate}

The attempt of measuring $s - s_{\rm FP}$ by the second sound will not
become obsolete by a possible detection of the sixth sound. The
reason is that the two modes are sensitive to different modifications
of the two-fluid equations. The sixth sound velocity is due to the
$s_{\rm s}$-contribution in the entropy continuity equation (28); the
$\mu_{\rm s}$-term in (30) yields higher order corrections to $c_6$
only. On the other hand, a main correction in the second sound
velocity stems from the $\mu_{\rm s}$-term in (30).

\section{Observability of the sixth sound}
We start this section with a simplified, alternative derivation of the
sixth sound. We discuss then in detail the possible observation of the
sixth sound.

\subsection{Entropy transport by the sixth sound}
We assume that the ring of Fig. 2 carries a persistent current with the
velocity $u_{\rm s}$, and that at a certain time there is the following
temperature and pressure variation along the ring:
\begin{equation}
\delta T (\phi ) = A\, \cos (n \hh \phi)\,,\qquad
\delta P(\phi) \approx \frac{s}{v}\; \delta T \,,
\qquad n\in\{1,2,\ldots\}\,.
\end{equation}
This variation implies $\mu \approx \mbox{const.}$ (the small l.h.s. of
(30) and the $\mu_{\rm s}$-term may be neglected for the present
purpose). The variation (75) is a fountain pressure gradient; for
$s_{\rm s} =0$ it is metastable.

Eq. (75) implies a corresponding variation of the entropy density:
\begin{equation}
\rho\,\h \delta s = \frac{\rho\, c_\mu }{T}\; \delta T (\phi) \,.
\end{equation}
Due to (68) we may use $\rho\approx \mbox{const.}$. The continuity
equation (49) implies then $\rho_{\rm s}\h u_{\rm s}\approx
\mbox{const.}$.

For $s_{\rm s}\ne 0$ the persistent current carries entropy. For
constant $T$ and $P$ the net entropy current $\delta j_s$ vanishes. For
(75) we obtain (using $\rho_{\rm s}\h u_{\rm s}\approx \mbox{const.}$):
\begin{equation}
\delta j_s = \rho_{\rm s}\, u_{\rm s}\,\h \delta s_{\rm s} = \rho_{\rm
s}\, u_{\rm s}\;\frac{ c_{\mu,\rm s}}{T}\; \delta T (\phi) \,.
\end{equation}
Due to this heat current the entropy variation (76) is shifted along
the ring. This shift takes place with the velocity
\begin{equation}
c_6 = \frac{\delta j_{\rm s}}{\rho\,\h \delta s} = \, u_{{\rm s}} \;
\frac{\rho_{\rm s}}{\rho }\; \frac{c_{\mu , \hh\rm s}}{c_{\mu }}\,.
\end{equation}
In this rather simple way the sixth sound may be understood as an
entropy transport phenomenon. Section 3 shows that it is also -- in
accordance with the common nomenclature -- a sound mode.

The net entropy current (77) must change the entropy variation of (76)
of the {\em whole ring}\/ (consisting of the container, the powder and
He$\;$II). For the considered experiment, the entropy density $\rho\h
s$ in (50) must therefore include the entropy of the powder and of the
container. In practice, this can be taken into account by equating
$c_\mu$ in (78) with the specific heat of the whole ring, $c_\mu =
c_{{\rm ring}}$. At the considered low temperatures, the specific heat
(per atom) of helium is at least four orders of magnitude larger than
that of normal solid material. Therefore, we may set $c_{{\rm
ring}}\approx c({\rm He})$. In addition, the specific heat at constant
chemical potential may be approximated by that at constant pressure
(implying an error of the order $10^{-2}$):
\begin{equation}
c_\mu = c_{{\rm ring}}\approx c_P({\rm He})\,,
\qquad \kappa = \kappa_{\rm ring}\approx \kappa({\rm He})\,.
\end{equation}
These arguments apply similarily for the heat conductivity $\kappa$.

\subsection{Shift of the temperature variation}
Any temperature and pressure variation in an actual ring experiment is
likely to have a FP component. In the presence of a persistent current
this variation is shifted with the velocity $c_6$ along the ring. This
shift should be observable. It has characteristic properties:
\begin{itemize}
\item
The shift velocity is proportional to that of the persistent current.
\item
The shift velocity is independent of the amplitude of the temperature
variation (as long as the linear approximation works).
\end{itemize}
The observation of the first property would leave no doubt that the
persistent current transports entropy, implying that the superfluid
entropy is non-zero.  The experiment would then determine this
superfluid entropy quantitatively.  If no such shift is observed the
experiment would yield an upper limit for $S_{\rm s}$ which is
considerably lower than the present one.

\begin{figure}[t]
\begin{center} \setlength{\unitlength}{9mm}
\begin{picture}(14,10.4)(0,-0.5)
\iffinalplot
\put(0.55,3.6){\line(1,0){12.85}}
\small
\put(-.13,3.6){\beginpicture
\setcoordinatesystem units <4.86cm,1.215cm>
\setplotarea x from 0 to 2.4, y from -2.4 to 5.8
\axis bottom ticks in withvalues 0 0.05 0.1 0.15 0.2 /
at 0 0.5 1 1.5 2 / /
\axis left ticks in numbered from -2 to 5 by 1 /
\axis right ticks in from -2 to 5 by 1 /
\axis top ticks in withvalues 2.1 2.0 1.9 1.8 1.7 /
at 0.3315 0.7919 1.2538 1.7157 2.1776 / /
\setquadratic
\setplotsymbol ({\viiipt\rm .})
 \plot 0.001   0.000
   0.010  -0.249     0.030  -0.740     0.050  -1.040     0.070  -1.257
   0.090  -1.416     0.110  -1.529     0.130  -1.607     0.150  -1.656
   0.170  -1.680     0.190  -1.682     0.210  -1.667     0.230  -1.637
   0.250  -1.593     0.270  -1.537     0.290  -1.471     0.310  -1.396
   0.330  -1.313     0.350  -1.223     0.370  -1.126     0.390  -1.025
   0.420  -0.864     0.460  -0.635     0.500  -0.395     0.540  -0.145
   0.580   0.110     0.620   0.369     0.660   0.631     0.700   0.893
   0.740   1.154     0.780   1.413     0.820   1.669     0.860   1.921
   0.900   2.167     0.940   2.408     0.980   2.643     1.020   2.870
   1.060   3.090     1.100   3.301     1.140   3.504     1.180   3.697
   1.220   3.881     1.260   4.055     1.300   4.218     1.340   4.370
   1.380   4.510     1.420   4.639     1.460   4.755     1.500   4.859
   1.540   4.950     1.580   5.027     1.620   5.091     1.660   5.141
   1.700   5.176     1.740   5.198     1.780   5.204     1.820   5.196
   1.860   5.172     1.900   5.133     1.940   5.078     1.980   5.007
   2.020   4.921     2.060   4.818     2.100   4.700     2.140   4.567
   2.180   4.417     2.220   4.253     2.260   4.074     2.300   3.880
   2.340   3.672     2.400   3.334 /
\endpicture }
\normalsize
\put(12.5,-.25){$|t|$}
\put(10.6,11.75){$T({\rm K})$}
\put(2,9.5){\large $\displaystyle \frac{c_6}{u_{\rm s}}$}
\put(-0.05,9.5){{\%}}
\fi
\end{picture} \end{center}
{\bf Figure 3}: Velocity $c_6$ of the sixth sound in the ring experiment of
Figure 2 as a function of the temperature $T$ for saturated vapour
pressure.
\end{figure}

For a potential experiment it is useful to know the expected size of
the effect.  We use $c_{\mu ,\hh {\rm s}}\approx c_{P,\hh {\rm
s}}$ which implies an error of the order $10^{-2}$. We insert this and
(79) into (78):
\begin{equation}
\frac{c_6}{u_{\rm s}} \approx \frac{\rho_{\rm s}}{\rho}
\;\frac{ T }{c_P} \;\frac{ \partial s_{\rm s}(T,P)}{\partial T
} \;.
\end{equation}
We use the prediction $s_{\rm s}(T,P)$ of Fig. 1 and the experimental
$\rho_{\rm s}/\rho$ and $c_P$. Fig. 3 displays the resulting
temperature dependence of $c_6/u_{\rm s}$.

The shift velocity of the temperature variation (75) along the ring
amounts up to a few percent of the supervelocity. Depending on the
temperature range the shift is parallel or antiparallel to the
persistent current (Fig. 3).

In an experiment, the temperature will be monitored at fixed points of
the ring. At a given point the temperature amplitude oscillates with
the frequency
\begin{equation}
\omega_{\rm FP} = \frac{c_6}{R}\;n\,.
\end{equation}
For $s_{\rm s} = 0$ (or $u_{\rm s} = 0$) the considered mode becomes
a static FP gradient along the ring. Therefore, we call (81) the
`fountain pressure' frequency.

As an example we insert the values $u_{\rm s} = 2\,{\rm cm/s}$ and $R =
2\,{\rm cm}$ of an actual experiment \cite{cl72}. Using $c_6/u_{\rm s}
\sim 3\%$ and $n=1$ we obtain an oscillation period $2\pi /\omega_{\rm
FP}$ of about three minutes.

Due to the damping (74) of the sixth sound the amplitude of the FP
oscillation will be reduced by a factor e after
\begin{equation}
n_{\rm osc} = \frac{\left|\h \omega_{\rm FP}\right| }{2\pi\hh
{\Gamma}_{\rm FP}} = \frac{T \rho_{\rm s}}{2\pi\h m\h \kappa
}\;\left| \, \frac{\partial s_{\rm s}(T,P)}{\partial T} \; u_{\rm s}\,
\right| \frac{R}{n}
\end{equation}
cycles of oscillation. We insert $n=1$, $R=2\,{\rm cm}$ and $u_{\rm
s}=2\,{\rm cm/s}$. We use the prediction $s_{\rm s}(T,P)$ of Fig. 1 and
the experimental $\rho_{\rm s}$. The heat conductivity $\kappa =
\kappa ({\rm He}) $ of He$\,$II is not well-known. For evaluating (82)
we assume $\kappa =0.05\,{\rm J/(m\, s\, K)}$; this value is cited in
\cite{pu74} as an upper limit for $T=2.1\,$K. The result is displayed
in Fig. 4.

\begin{figure}[t]
\begin{center} \setlength{\unitlength}{9mm}
\begin{picture}(14,5.5)(0,-0.5)
\iffinalplot
\small
\put(-.13,0){\beginpicture
\setquadratic
\setcoordinatesystem units <4.86cm,0.135cm>
\setplotarea x from 0 to 2.38, y from 0 to 38
\axis bottom ticks in withvalues 0 0.05 0.1 0.15 0.2 /
at 0 0.5 1 1.5 2 / /
\axis left ticks in numbered from 10 to 30 by 10 /
\axis right ticks in from 10 to 30 by 10 /
\axis top ticks in withvalues 2.1 2.0 1.9 1.8 1.7 /
at 0.3315 0.7919 1.2538 1.7157 2.1776 / /
\setplotsymbol ({\viiipt\rm .})
 \plot 0.000   0.000
  0.010  5.766   0.030 14.974   0.050 19.588   0.070 22.463
  0.090 24.243   0.110 25.257   0.130 25.705   0.150 25.721
  0.170 25.401   0.190 24.817   0.210 24.023   0.230 23.062
  0.250 21.966   0.270 20.766   0.290 19.483   0.310 18.135
  0.330 16.739   0.350 15.308   0.370 13.851   0.390 12.382
  0.420 10.178   0.460  7.239   0.500  4.354   0.5634 0 /
\plot 0.5634 0
  0.580  1.139   0.620  3.708   0.660  6.143   0.700  8.434
  0.740 10.577   0.780 12.567   0.820 14.405   0.860 16.090
  0.900 17.623   0.940 19.005   0.980 20.241   1.020 21.335
  1.060 22.288   1.100 23.107   1.140 23.795   1.180 24.357
  1.220 24.799   1.260 25.125   1.300 25.340   1.340 25.451
  1.380 25.462   1.420 25.379   1.460 25.207   1.500 24.951
  1.540 24.617   1.580 24.210   1.620 23.735   1.660 23.198
  1.700 22.598   1.740 21.956   1.780 21.261   1.820 20.525
  1.860 19.750   1.900 18.943   1.940 18.107   1.980 17.248
  2.020 16.370   2.060 15.476   2.100 14.572   2.140 13.663
  2.180 12.750   2.220 11.840   2.260 10.935   2.300 10.040
  2.340  9.158   2.380  8.292 /
\endpicture }
\normalsize
\put(12.2,-0.55){$|t|$}
\put(10.5,5.95){$T({\rm K})$}
\put(1.8,4){\large $n_{\rm osc}$}
\fi
\end{picture} \end{center}
{\bf Figure 4}: Number $n_{\rm osc}$ of observable fountain pressure
oscillations for a ring experiment ($R=2\,{\rm cm}$, $u_s = 2\,{\rm
cm/s})$ as a function of the temperature at saturated vapour
pressure.
\end{figure}

At the maximum of $s_{\rm s}$ in Fig. 1 the velocity $c_6$ and the
number $n_{\rm osc}$ vanish. The exact position of this point is
subject to the uncertainty in the predicted $s_{\rm s}$. It is also
slightly shifted by the approximation $c_{\mu ,{\rm s}}\approx
c_{P,{\rm s}}$.

The condition $n_{\rm osc}\gg 1$ for easy observability is fulfilled at
most temperatures. Moreover, the number $n_{\rm osc}\propto |u_{\rm
s}|\h R$ of observable oscillations could be increased by using a
higher supervelocity or a larger ring. The sixth sound or,
equivalently, the FP oscillations should be readily detectable.

The integration of eq. (80) yields
\begin{equation}
s_{\rm s}(T,P) \approx \int_{T_\lambda}^T \! dT'\;
\frac{c_6/u_{\rm s} } {\rho_{\rm s} /\rho}
\; \frac{ c_P }{T'}\;.
\end{equation}
The quantities $c_6/u_{\rm s}$, $\rho_{\rm s} /\rho$ and $c_P$ are
functions of $T'$ and $P$. All these quantities can be measured.  For
the evaluation of the integral they have to be measured in the
temperature range from $T_\lambda$ (where $c_6/u_{\rm s}$ and
$\rho_{\rm s} /\rho$ vanish) to $T$ and at fixed pressure. In this way
the proposed experiment determines the superfluid entropy $s_{\rm
s}(T,P)$.

If the sixth sound is not observed the considered experiment yields an
upper limit for the superfluid entropy which is roughly given by
\begin{equation}
\left(\frac {S_{\rm s}} {S}\right)_{\!{\rm upper~limit}}
\approx
\; \frac{1}{n_{\rm osc}} \, \left( \frac {S_{\rm s}} {S}
\right)_{\!{\rm predicted}} \sim \; \frac{1\% }{n_{\rm osc}}\;.
\end{equation}
In this way the present upper limit could be lowered by at least one
order of magnitude.

\section{Concluding remarks}

A superfluid entropy of relative size $S_{\rm s}/S\sim 1\%$ is not
excluded experimentally. We have presented a straight-forward
generalization of the two-fluid equations for a non-vanishing
superfluid entropy. The investigation of sound modes leads to a
specific proposal by which a superfluid entropy of the considered size
could be detected. A negative experiment would yield a new upper limit
for the entropy content of the superfluid fraction.

\end{document}